\documentclass[11pt]{article}

\usepackage{epsf}
\usepackage{amsmath,amssymb}
\usepackage[dvips]{color}
\usepackage{graphicx}
\usepackage{epsfig}
\newcommand{\be}{\begin{equation}}
\newcommand{\ee}{\end{equation}}
\newcommand{\bea}{\begin{eqnarray}}
\newcommand{\eea}{\end{eqnarray}}
\newcommand{\ba}{\begin{array}}
\newcommand{\ea}{\end{array}}
\def \nn {\nonumber}
\numberwithin{equation}{section}

\def\bid{\hbox{1\hspace{-0.04in}I}} %blackboard bold 1
\def\bbz{\hbox{O\hspace{-0.10in}0}} %blackboard bold 0
\def\Z {\mathbb{Z}}
\def\R {\mathbb{R}}

\begin{document}
%%%%%%%%%%%%%%%%%%%%%%%%%%%%%%%%

%%%%%%%%%%%%%%%%%%%%%%%%%%
\begin{titlepage}

\vfill

\begin{center}

{\hfill 2011/12/07}

\bigskip

{\Large \bf{Double Field Theory for Double D-branes}}

\vfill

Cecilia Albertsson$^a$\footnote{\tt  cecilia.albertsson@gmail.com}, Shou-Huang Dai$^b$\footnote{\tt  shdai@ntnu.edu.tw}, Pei-Wen Kao$^c$\footnote{\tt kao06@math.keio.ac.jp}, Feng-Li Lin$^b$\footnote{\tt linfengli@phy.ntnu.edu.tw}
\bigskip

{\it $^a$ Lunascape Corporation, Sotokanda 1-18-13, Tokyo 101-0021, Japan \\}
{\it $^b$ Department of Physics, National Taiwan Normal University, \\
88, Sec.4, Ting-Chou Road, Taipei 11677, Taiwan\\}
{\it $^c$ Department of Mathematics, Keio University, 3-14-1, Hiyoshi, \\Kohoku-ku, 223-8522 Yokohama, Japan\\}

\end{center}

\vfill
\begin{abstract}

 We consider Hull's doubled formalism for open strings on D-branes in flat space and construct the corresponding effective double field theory. We show that the worldsheet boundary conditions of the doubled formalism describe in a unified way a T-dual pair of D-branes, which we call double D-branes. We evaluate the one-loop beta function for the boundary gauge coupling and then obtain the effective field theory for the double D-branes.  The effective field theory is described by a DBI action of double fields. The T-duality covariant form of this DBI action is thus a kind of ``master'' action, which describes all the double D-brane configurations related by T-duality transformations. We discuss a number of aspects of this effective theory.

\end{abstract}
\vfill

\end{titlepage}

%%%%%%%% BODY %%%%%%%%%%%%%
\setcounter{footnote}{0}
%%%%%%%%%%%%%%%%%%%%%%%%%%%%%%%

\section{Introduction}
 T-duality \cite{Kikkawa:1984cp,Buscher:1987sk,Buscher:1987qj,Giveon:1994fu} is one of the most important symmetries in string theory. T-duality transformations relate different types of string theories, including D-branes on distinct geometries. They moreover exchange the roles of the metric and the NS $B$-field, as well as the roles of momentum and winding modes. It was recently discovered that T-duality transformations can yield non-geometric backgrounds \cite{Dabholkar:2002sy,Kachru:2002sk,Hellerman:2002ax,Flournoy:2004vn,Schulz:2011ye}, a fact which enlarges the range of allowed background spacetimes in string theory. One such non-geometric target space is the T-fold, on which T-duality transformations play the role of transition functions to bridge different patches. A novel proposal by Hull \cite{Hull:2004in,Dabholkar:2005ve,Hull:2006qs,Hull:2006va} is to treat such non-geometric backgrounds in a kind of embedding geometry, so-called doubled geometry. Doubled geometry may be constructed by duplicating the target space dimensions in such a way as to put the original space and its T-dual on the same footing. In the case of T-folds the non-geometric T-duality transition functions are lifted to geometric ones on the doubled space, rendering the doubled geometry covariant under T-duality transformations. In this way a T-duality symmetric formulation of string theory allows for non-geometric target spaces. The need for a T-duality symmetric description explains why T-folds were not recognized as relevant earlier, as there was no natural way of incorporating them as backgrounds.

Until the recent work of Hull et al, the search for theories with manifest T-duality symmetry was not given excessive attention, but a few efforts stand out. The T-duality symmetric formulation proposed by Tseytlin\footnote{For earlier work, see for example \cite{Duff:1986ne,Duff:1989tf}.} \cite{Tseytlin:1990va,Tseytlin:1990nb}, and subsequently discussed in \cite{Maharana:1992my,Schwarz:1993vs} consists in rewriting the standard worldsheet theory as a doubled theory by adding the dual coordinates to the target space.  To avoid doubling the degrees of freedom by the addition of coordinates, the target space coordinates are treated as chiral fields with a worldsheet action of the form proposed by Floreanini and Jackiw (FJ) \cite{Floreanini:1987as}. We thus end up with an FJ-type action which is not manifestly Lorentz covariant. To ensure on-shell worldsheet covariance, the doubled geometry of the target space is required to be T-duality covariant \cite{Tseytlin:1990va}.  This was the first time that a T-duality symmetric string theory had been formulated. Hull reformulated this doubled framework in an alternative way by instead having the worldsheet Lorentz symmetry manifest, and eliminating half of the target space degrees of freedom by way of a self-duality constraint \cite{Hull:2004in,Hull:2006va}. The target space metric in the worldsheet action of Hull's doubled formalism is the metric on the doubled geometry.   It turns out that Tseytlin's and Hull's formulations are equivalent; this was shown in \cite{Berman:2007xn} by way of gauging the Pasti-Sorokin-Tonin (PST) formulation \cite{PST} of Hull's doubled formalism.

       The doubled theory is equivalent to the standard worldsheet theory at the on-shell level once the self-duality condition is imposed. One may expect the equivalence to hold also at the quantum level. One way to verify this is to evaluate the one-loop beta functions of the doubled formalism and compare them with the ones from the standard worldsheet formalism, while imposing the self-duality constraint. This check was done in \cite{Berman:2007xn}, showing that quantum equivalence indeed holds.  An important by-product of this work was the  effective double field theory for gravity, from which the equations of motion can be derived by setting the one-loop beta functions to zero, before imposing the self-duality condition.  Although this double field theory is not intrinsically different from the usual gravity theory obtained from string theory in the low energy limit, it is new from the doubled geometry perspective.

Recently, an alternative approach to obtaining an intrinsic double field theory for the massless sector of closed strings has been developed in a series of papers by Hull, Zwiebach and their collaborators \cite{Hull:2009mi,Hull:2009zb,Hohm:2010jy,Hohm:2010pp,Hohm:2010xe}, see also earlier work by Siegel \cite{Siegel:1993th}.\footnote{In \cite{Thompson:2011uw}, Thompson shows that the duality invariant approach to M-theory in \cite{Berman:2010is, Berman:2011pe} is related to the double field theory of \cite{Hull:2009mi}$-$\cite{Hohm:2010pp} by the doubled Kaluza-Klein reduction.} This approach is based on a T-duality symmetric closed string field theory \cite{Kugo:1992md}, from which the gauge algebra of the double field theory for the massless string fields can be constructed systematically. The gauge algebra turns out to be a (deformed) Courant algebra \cite{Hull:2009zb}, from which a gauge invariant double field theory for the massless closed string sector can be constructed based on symmetry principles \cite{Hohm:2010jy,Hohm:2010pp}. Although the original closed string theory of this approach is intrinsically doubled, a strong constraint from the level matching condition must be imposed on the string fields and their products in order to arrive at a background independent action. This strong constraint then kills half of the doubled degrees of freedom and brings the action to an undoubled effective field theory of massless closed string fields.  An attempt was made recently in \cite{Copland:2011yh} to in a special case show the equivalence between double field theories constructed via the two approaches above. In addition, a unified description of the low-energy limits of type II string theories with respect to T-duality was proposed in \cite{Hohm:2011zr,Hohm:2011dv}.

       Inspired by the double field theory constructions for the massless closed string sector, we derive in this paper the effective double field theory for the massless open string sector on D-branes in Hull's doubled geometry formalism. Unlike earlier studies of D-branes in the doubled formalism \cite{Lawrence:2006ma,Albertsson:2008gq}, which formulated the consistency conditions and found the nontrivial D-brane embeddings in doubled geometry, we will instead consider  D-branes in a flat space with a constant $B$-field.  We show explicitly that the boundary conditions in the doubled formalism define T-dual pairs of D-branes, i.e., double D-branes. Furthermore, we show that there is a worldsheet doubled action with a T-duality covariant boundary gauge coupling. Finally, we evaluate the effective double field theory for the double D-branes, after using a generalization to doubled formalism of the usual background field method to derive the DBI action \cite{Abouelsaood:1986gd,Mukhi:1985vy}. We find that our action is of the DBI form for the double fields on the worldvolume of double D-branes. Rewriting this action on the T-duality covariant form, it becomes what we refer to as a master action for all the double D-brane configurations related by T-duality transformations. This is the first time that such an action has been derived. Moreover,  after applying the self-duality condition, the effective action can be reduced to the usual DBI action for a single D-brane. However,  the relation between bulk and boundary gauge symmetries is not clear in our setup. Our $B$-field dependence is different from the one expected for the usual DBI action; this is an interesting issue which merits further investigation.

  This paper is organized as follows. Section~\ref{SecDGreview} begins by reviewing Hull's doubled formalism worldsheet action, followed by a summary of the three conditions relevant to us, which the Neumann and Dirichlet projectors that define our D-branes must satisfy on doubled geometry. Explicit examples are presented to show that a unified description for T-dual pairs of D-branes is encoded in the doubled formalism. For the purpose of quantizing the double field action, which comes with a self-duality constraint at the level of equations of motion, the FJ-type action is introduced in section~\ref{SecFJ}, and its relation to the PST action is explained. Then we propose a boundary gauge coupling term for the worldsheet doubled action, and derive the effective theory by means of the background field method. Several issues concerning this effective theory are discussed. Section~\ref{SecConclusion} contains our conclusions, and some technical details are collected in the appendices. Appendix \ref{AppOnn} describes how to solve for the generic boundary projectors in the chiral frame as well as in the light-like $O(n,n)$ frame. Appendix \ref{AppO22} focuses on the $O(2,2)$ case, for which we list the explicit Neumann and Dirichlet projectors allowed by the doubled geometry properties. Appendix \ref{AppGreen} provides the details of the calculation of the Neumann Green's function.

\section{Doubled formalism of open strings}
\label{SecDGreview}

  In this section we review the doubled formalism for string theory. Originally the doubled formalism \cite{Hull:2004in,Hull:2006va} was used to describe string theory on a target space that is a T-fold, namely,  locally a $\mathbb{T}^{n}$-bundle such that the transition functions are taken from the T-duality group $O(n,n;\Z)$.  The purpose of this formalism is to make the T-duality manifest as a symmetry at the level of the worldsheet action, and thereby obtain a geometric description of the non-geometric T-fold. This is achieved by doubling the target space coordinates but at the same time paying the price of imposing a self-duality constraint. Many interesting results  based on this formalism have been discussed, see for example the general discussions in \cite{Hull:2009sg}.  On the other hand, it seems quite trivial to apply the doubled formalism to flat space without the subtle obstruction of a torus fibration. This is true if we consider the closed string theory. However, for open string theory with D-brane embedding, we will see some interesting results even in flat space. Moreover, we can then derive the effective double field theory for D-branes in this framework.

 \subsection{Basics of doubled formalism}

    In this paper we consider the D-brane in flat spacetime, so we choose a very simple doubled geometry, the doubled flat space. Then the doubled formalism just follows as given in \cite{Hull:2004in,Hull:2006va}. Our doubled space is locally a $2n$-dimensional flat space labeled by $2n$ coordinates $\{ \mathbb{X}^{I}| I=1,\cdots, 2n \}$  fibered trivially over a time-like coordinate\footnote{In general, one can also consider time-like T-duality as in \cite{Hull:1998br,Hull:1998vg}, and then introduce the corresponding double coordinate. However, to avoid complication we restrict ourselves to only space-like doubled space.} $T$. The  string theory on this target space is described by the worldsheet action
\begin{equation} \label{doubledaction}
S_{w.s.}=\int d^2\sigma \left( -\frac{1}{2}\mathbb{H}_{IJ}\eta^{\alpha\beta}\partial_{\alpha}\mathbb{X}^{I}\partial_{\beta}\mathbb{X}^{J} + \frac{1}{2}\eta^{\alpha\beta}\partial_{\alpha}T\partial_{\beta}T \right) \;.
\end{equation}
Here $\eta=\mathrm{diag}(-1,1)$ is the flat worldsheet metric, and $\mathbb{H}$ is the metric on the doubled geometry target space. In this paper, we only consider the doubled flat space for which $\mathbb{H}$ is a constant $2n \times 2n$ matrix. Note that the $T$-coordinate part in (\ref{doubledaction}) has time-like signature.  The action (\ref{doubledaction}) is invariant under $O(n,n)$ (T-duality) transformations: for $\mathit{h} \in O(n,n)$, the double coordinates and the doubled flat space metric transform as (the superscript $t$ denotes transpose)
\be
\mathbb{X} \longrightarrow \mathit{h}^{-1}\; \mathbb{X}\;, \qquad  \mathbb{H} \longrightarrow \mathit{h}^t\; \mathbb{H} \; \mathit{h}.
 \ee
We also define an $O(n,n)$-invariant metric $\mathbb{L}_{IJ}$ for the tangent space of doubled geometry, i.e.,
\be
ds_{O(n,n)}^2=\mathbb{L}_{IJ} d\mathbb{X}^I d\mathbb{X}^J \qquad \mbox{with}\qquad   \mathit{h}^t \; \mathbb{L}\; \mathit{h}=\mathbb{L},
\ee
which is used to raise and lower the doubled space indices.\footnote{Note that we use a non-standard notation for the metrics $\mathbb{H}_{IJ}$ and $\mathbb{L}_{IJ}$; these are usually denoted in the literature by $\mathcal{H}_{IJ}$ and $L_{IJ}$, respectively.}

The bulk equations of motion generated by (\ref{doubledaction}) are
\be
\eta^{\alpha\beta}\; \mathbb{H}_{IJ}\; \partial_{\alpha}\partial_{\beta} \mathbb{X}^J=0\;, \qquad \eta^{\alpha\beta} \partial_{\alpha}\; \partial_{\beta} T=0\;.
\ee
Besides the equations of motion, we need to impose the self-duality condition \cite{Hull:2004in}
\begin{equation}\label{selfdualcondition}
\partial_{\alpha}\mathbb{X}^I=\epsilon_{\alpha \beta}\, \mathbb{L}^{IJ}\mathbb{H}_{JK}\,\partial^{\beta}\mathbb{X}^K
\end{equation}
to eliminate half of the degrees of freedom and reduce to the usual non-doubled worldsheet description.
Note that the self-duality condition \eqref{selfdualcondition} is $O(n,n)$ invariant but not $GL(2n)$ invariant,
so that the manifest $GL(2n)$ symmetry of the worldsheet theory \eqref{doubledaction} is broken to $O(n,n)$
by the self-duality constraint. However, the doubled theory is still  $GL(2n)$ covariant, and we can use this
covariance to change the frame by choosing a different $O(n,n)$ invariant metric.

   Thanks to the $O(n,n)$ symmetry, we can choose a polarization, i.e., a particular frame, to decompose the double coordinates of the $O(n,n)$ representation to obtain a $GL(n)\oplus GL(n)$ representation. This decomposes the doubled space into a T-dual pair of spaces. Then,
\be \label{X}
\mathbb{X}^I=(X^i,\tilde{X}_i)^t\;,
\ee
where $\{X^i | i=1,\cdots, n \}$ and $\{\tilde{X}_i | i=1,\cdots, n \}$ are the respective coordinates on each of the flat spaces in the T-dual pair.  The $O(n,n)$ invariant metric takes a light-like form in this frame, i.e., $ds_{O(n,n)}^2=2dX^i d\tilde{X}_i$ and
\be \label{L}
\mathbb{L}=\left( {\begin{array}{cc}
 \bbz_{n \times n} & \bid_{n \times n}  \\
 \bid_{n \times n} &  \bbz_{n \times n}  \\
 \end{array} } \right)\;,
\ee
where $\bid$ and $\bbz$ represent the identity matrix and the matrix of zeros, respectively.
The doubled space metric can be written on an $O(n,n)/O(n)\times O(n)$ coset form as
\be\label{Hmetric}
\mathbb{H}= \left(\begin{array}{cc}
g_{ij}-B_{ik}g^{kl}B_{lj} & B_{ik}g^{kj} \\
-g^{ik}B_{kj} & g^{ij} \\
\end{array}\right) \;, \qquad i,j,k,l=1,\cdots,n\;.
\ee
Here the symmetric field $g$ and the antisymmetric field $B$ are a flat space metric and a constant NS 2-form, respectively, on the space $\mathbb{T}^n$ or $\R^n$ spanned by the coordinates\footnote{In this paper, we are mainly interested in deriving the effective double field theory for zero modes, so we ignore the compactness of the flat target space, as well as the difference between $O(n,n;\Z)$ and $O(n,n;\R)$.} $\{X^i\}$.

\subsection{D-brane embedding in doubled flat space}

    It is well-known that if a target space coordinate of an open string satisfies the Dirichlet boundary condition, then its T-dual satisfies the Neumann boundary condition, and vice versa. Therefore, in the doubled formalism, half of the components in $\{\mathbb{X}^I\}$ obey the Dirichlet condition, and the other half, being the dual coordinates, obey the Neumann condition. Thus a D-brane and its T-dual can be described simultaneously in doubled formalism.  Although this aspect of doubled formalism as a unified description of D-branes may seem obvious, it has not been greatly emphasized in the literature. In this section we review the basics of D-brane embedding in doubled formalism along the lines of \cite{Lawrence:2006ma,Albertsson:2008gq} and demonstrate the unified description explicitly.

    In doubled formalism, the D-brane embedding may be defined by constructing a Dirichlet projector $\Pi^I_{D,J}$ which projects vectors onto the Dirichlet directions (i.e., the directions normal to the brane) in doubled space. We can then define the corresponding Neumann (i.e., tangent) projector as $\Pi_{N,J}^{\phantom{N,J}I}\equiv (\bid-\Pi^t_D)_J^{\ I}$ . By definition these complementary projectors are idempotent,
\be\label{projc}
\Pi^2_D=\Pi_D\;, \qquad \Pi^2_N=\Pi_N\;.
\ee

The Dirichlet projector is used to express the Dirichlet boundary conditions in a covariant way: the derivative of the Dirichlet target space coordinates with respect to the time-like worldsheet parameter $\sigma^0$ must vanish:
\be\label{Dcond}
\Pi^I_{D,J} \partial_0 \mathbb{X}^J|_{\partial_\Sigma}=0\;,
\ee
where $\partial \Sigma$ denotes the boundary of the worldsheet $\Sigma$.

Inserting the Dirichlet condition (\ref{Dcond}) into the boundary equations of motion derived by varying the action (\ref{doubledaction}) with respect to the doubled target space coordinates, yields the Neumann boundary condition,
\be\label{Ncond1}
0=\delta \mathbb{X}^I \mathbb{H}_{IJ} \partial_1 \mathbb{X}^J|_{\partial \Sigma}=\delta\mathbb{X}^I \Pi_{N,I}^{\phantom{N,I}J}\mathbb{H}_{JK} \partial_1 \mathbb{X}^K|_{\partial \Sigma}\;,
\ee
or equivalently,
\be\label{Ncond2}
\Pi_{N,I}^{\phantom{N,I}J}\mathbb{H}_{JK} \partial_1 \mathbb{X}^K|_{\partial \Sigma}=0\;.
\ee
To arrive at the second equality in (\ref{Ncond1}), we have used the fact that $\bid=\Pi_N+\Pi^t_D$ and $\delta \mathbb{X}\; \Pi^t_D=0$ (the latter is equivalent to the condition (\ref{Dcond})).

Because the doubled formalism is T-duality invariant, the Dirichlet condition (\ref{Dcond}) and the Neumann condition (\ref{Ncond2}) may be said to be equivalent; this follows immediately from the self-duality condition (\ref{selfdualcondition}), which defines a relation between the mutually dual coordinates $\{X^i\}$ and $\{\tilde{X}_i\}$. However, this statement presupposes the consistency of the Dirichlet and Neumann projectors that we have defined. First we need to ensure that they are compatible with the properties of doubled geometry.

   The Neumann condition (\ref{Ncond2}) is consistent with the self-duality condition (\ref{selfdualcondition}) only if
\be
\Pi_{N,I}^{\phantom{N,I}J} \mathbb{L}_{JK}\partial_0 \mathbb{X}^K|_{\partial\Sigma}=0.
\ee
The projector property (\ref{projc}) reveals that this condition is in fact equivalent to a null condition for both projectors,
\be\label{nullc}
\Pi^t_D \; \mathbb{L}\; \Pi_D=0\;, \qquad \Pi_N \; \mathbb{L} \; \Pi^t_N=0.
\ee
Note that these two conditions are not identical, in that they provide different statements about the D-brane embedding.

In addition to the conditions (\ref{nullc}), which are required for consistency of the boundary conditions, we also demand that the Neumann and Dirichlet projectors be mutually orthogonal with respect to the doubled space metric, on physical grounds: vectors tangent and normal to the brane should be locally orthogonal with respect to the metric on the relevant space. We thus have the orthogonality condition
\be\label{orthocon}
\Pi_N\; \mathbb{H}\; \Pi_D=0\;.
\ee
Note that in (\ref{Dcond}) the Dirichlet directions on the open string boundary are manifestly uniquely determined by the projector $\Pi_D$, whereas the Neumann directions in (\ref{Ncond2}) seem to mix with the Dirichlet ones in that the contraction $\mathbb{H}_{JK} \partial_1 \mathbb{X}^K$ runs over all $K$. However, under the requirement of orthogonality (\ref{orthocon}), this is actually not true, and the Neumann boundary condition only involves contraction in the Neumann directions,
\be
 (\Pi_{N})_I{}^J \mathbb{H}_{JK} \partial_1 \mathbb{X}^K = (\Pi_{N})_I{}^J \mathbb{H}_{JK} (\Pi_{N}^t){}^K{}_L \partial_1 \mathbb{X}^L \ .
\ee
Hence also the Neumann directions are uniquely specified by the projector $\Pi_N$.

  Finally, as was shown in \cite{Albertsson:2008gq} for a more general setting with D-branes embedded in a generic doubled geometry, we need to impose the integrability condition
\be\label{integcon}
\Pi_{N,I}^{\phantom{N,I}I'}\Pi_{N,J}^{\phantom{N,J}J'}\partial_{[I'}\Pi_{N,J']}^{t \phantom{N,J']}K}=0\;,
\ee
to ensure that our D-brane is locally a smooth submanifold of the target space. However, because our target space is flat and the $B$-field constant, the projectors are coordinate independent and trivially integrable.\footnote{There was one more condition derived in \cite{Albertsson:2008gq}, related to a Wess-Zumino term, but since we have no such term in our setup, this condition may be safely ignored.}

   To summarize, the D-brane embeddings allowed in our simple flat doubled space may be deduced by solving the four conditions (\ref{projc}), (\ref{nullc}), (\ref{orthocon}) and (\ref{integcon}) for the explicit forms of the Neumann and Dirichlet projectors. The conditions (\ref{projc}) and (\ref{nullc}) are necessary for the projectors to define the appropriate boundary conditions for D-branes, while the conditions  (\ref{orthocon}) and (\ref{integcon}) were motivated by the commonly adopted assumption that the theory on the doubled space must be physical in order to produce physical theories on its physical subspace components. A relatively simple way of computing the projectors is illustrated in Appendix~\ref{AppOnn}, and the most general solutions of $\Pi_D$ and $\Pi_N$ satisfying the above set of conditions for the $n=2$ case (i.e., when the doubled space is 2$n$ = 4-dimensional) are derived in detail in Appendix \ref{AppO22}. The solutions include D0-, D1- and D2-branes\footnote{Note that all branes in doubled geometry have the same dimension, namely half of the dimension of the full space; in the case considered here they are all two-dimensional. The labels D0, D1 and D2 refer to the number of dimensions a brane has in one of the $\mathbb{T}^2$-components of the doubled space.} compactified on our  $\mathbb{T}^4$.

\subsection{Examples}
\label{SecExamples}

Here we consider a few simple examples of projectors that solve the conditions (\ref{projc}) and (\ref{nullc}). We use these solutions to demonstrate the power of doubled formalism in providing a unified description for a T-dual pair of D-branes.

   Consider a 4-dimensional doubled flat space with constant $B$-field. If we define the light-like
 $O(2,2)$ metric as in (\ref{L}) on this space, the double coordinates may be split as
 \be \label{R4}
 \mathbb{X}=(X,Y,\tilde{X},\tilde{Y})^t\;.
 \ee
Now suppose there is a single D-brane living in the $\{X,Y\}$-space, plus its T-dual D-brane living in the $\{\tilde{X},\tilde{Y}\}$-space.

Because the target space is flat with a constant $B$-field, we have $g_{ij}=\delta_{ij}$ and $B_{ij}=\epsilon_{ij} B$ for $i,j=1,2$, with $B$ a constant real number. Then the doubled space metric (\ref{Hmetric}) reduces to
\begin{equation}
\mathbb{H}=\left( \begin{array}{cccc}
1+B^{2} & 0 & 0 & B  \\
0 & 1+B^2 & -B & 0\\
0 & -B & 1 & 0\\
B & 0 & 0 & 1
\end{array} \right),
\end{equation}
and the explicit components of the self-duality condition (\ref{selfdualcondition}) read
\begin{equation}
\label{selfdualex1}
\begin{array}{rcl}
\partial_{0}\tilde{X} &=& \partial_{1}X+B\partial_{0}Y \;,\\
\partial_{0}\tilde{Y} &=& \partial_{1}Y-B\partial_{0}X \;,\\
\partial_{1}\tilde{X} &=& \partial_{0}X+B\partial_{1}Y \;,\\
\partial_{1}\tilde{Y} &=& \partial_{0}Y-B\partial_{1}X\;,
\end{array}
\end{equation}
 or, if one prefers the inverse relations,
\begin{equation}
\label{selfdualex2}
\begin{array}{rcl}
\partial_{0} X &=&\frac{1}{1+B^2} (\partial_{1}\tilde{X}-B \partial_{0}\tilde{Y}) \;,\\
\partial_{0} Y &=& \frac{1}{1+B^2}( \partial_{1}\tilde{Y}+B\partial_{0}\tilde{X}) \;,\\
\partial_{1} X &=&\frac{1}{1+B^2} (\partial_{0}\tilde{X}-B \partial_{1}\tilde{Y})\;,\\
\partial_{1} Y &=& \frac{1}{1+B^2}( \partial_{0}\tilde{Y}+B\partial_{1}\tilde{X}) \;.
\end{array}
\end{equation}

 We now consider the following Dirichlet projector,
\begin{equation}
\Pi_D(D0)=\left( \begin{array}{cc}
 \bid_{2\times 2} &  \bbz_{2\times 2}   \\
 \bbz_{2\times 2} &   \bbz_{2\times 2}  \\
\end{array} \right).
\end{equation}
This projector is a solution of conditions (\ref{projc}) and (\ref{nullc})\footnote{However, it does not satisfy the orthogonality condition (\ref{orthocon}) unless $B=0$, as discussed in Appendix \ref{AppO22}.}, and inserting it into the Dirichlet condition (\ref{Dcond}) we see that it defines $X$ and $Y$ as Dirichlet directions.

Then the Neumann condition (\ref{Ncond2}) reduces to
\be\label{ex1}
(-B\partial_1 Y+\partial_1\tilde{X})|_{\partial \Sigma}=0\;, \qquad (B\partial_1 X+\partial_1 \tilde{Y})|_{\partial\Sigma}=0\;.
\ee
 At  first sight, this does not look like sensible Neumann or Dirichlet boundary conditions. However, after substituting (\ref{selfdualex1}) into  (\ref{ex1}), we obtain the boundary conditions for a D0-brane in the $\{X,Y\}$-space, i.e.,
 \be
 \partial_0 X|_{\partial \Sigma}=0\;, \qquad \partial_0Y|_{\partial\Sigma}=0\;,
 \ee
 which is nothing but the Dirichlet condition (\ref{Dcond}). In this sense, the Neumann condition (\ref{Ncond2})  and the Dirichlet condition (\ref{Dcond}) are equivalent, due to the presence of the self-duality condition.   If on the other hand we substitute (\ref{selfdualex2}) into (\ref{ex1}), we  arrive at the Neumann boundary conditions with $B$-field for $\tilde{X}$ and $\tilde{Y}$, namely,
 \be
 (\partial_1 \tilde{X} -B \partial_0 \tilde{Y})|_{\partial \Sigma}=0\;, \qquad  (\partial_1 \tilde{Y} +B\partial_0 \tilde{X})|_{\partial \Sigma}=0\;.
 \ee
 These boundary conditions describe a D2-brane in a constant $B$-field background in the $\{\tilde{X},\tilde{Y}\}$-space. This example shows that one can describe a T-dual pair of D0- and D2-branes through either the Neumann or the Dirichlet conditions when the self-duality condition is present.  Moreover, if we perform T-duality along both directions of either subspace ($\{X,Y\}$ or $\{\tilde{X},\tilde{Y}\}$), we see that the resultant new D-brane configurations are encoded in the Dirichlet projector
\begin{equation}
\Pi_D(D2)=\left( \begin{array}{cc}
 \bbz_{2\times 2} &  \bbz_{2\times 2}   \\
 \bbz_{2\times 2} &   \bid_{2\times 2}  \\
\end{array} \right).
\end{equation}

  Similarly, we can consider other Dirichlet projectors and arrive at analogous conclusions. For example, take the projectors for a pair of T-dual D-branes, such as\footnote{Note that $\Pi_{D}(D2) \equiv \Pi_{N}(D0)$ and $\Pi_{D}(D1(y)) \equiv \Pi_{N}(D1(x))$.}
 \bea\label{ProjEx2}
&&
\Pi_{D}(D1(x))=\left( \begin{array}{cccc}
1 & 0 & 0 & 0  \\
0 & 0 & 0 & 0  \\
0 & 0 & 0 & 0  \\
0 & 0 & 0 & 1
\end{array}\right) \ , \hspace{0.5cm}
\Pi_{D}(D1(y))=\left(\begin{array}{cccc}
0 & 0 & 0 & 0  \\
0 & 1 & 0 & 0  \\
0 & 0 & 1 & 0  \\
0 & 0 & 0 & 0
\end{array}\right) \;. \hspace{0.5cm}
 \label{pD1y}
\eea
Via the boundary conditions and the self-duality constraint as above, the projectors (\ref{ProjEx2}) will yield the expected dual pair of boundary conditions, namely,
 \bea
&\mbox{D1(x):}&\hspace{1cm}
\left\{\begin{array}{c}
\partial_0 X = 0 \\
\partial_1 Y = 0
\end{array}\right.
 \hspace{1cm} \mbox{or equiv.} \hspace{1cm}
\left\{\begin{array}{c}
\partial_1 \tilde{X} = 0 \\
\partial_0 \tilde{Y} = 0
\end{array}\right.  \label{bcD1x}\\
&\mbox{D1(y):}&\hspace{1cm}
\left\{\begin{array}{c}
\partial_1 X = 0 \\
\partial_0 Y = 0
\end{array}\right. \hspace{1cm} \mbox{or equiv.} \hspace{1cm}
\left\{\begin{array}{c}
\partial_0 \tilde{X} = 0 \\
\partial_1 \tilde{Y} = 0
\end{array}\right. \label{bcD1y}
\eea
Note that the constant $B$-field does not appear in the D1 boundary conditions since it cannot live on D1-branes.

\section{Effective double field theory for double D-branes}
\label{SecFJ}

   As shown in the previous section, the doubled formalism provides a T-duality symmetric description  of D-branes at the level of open string boundary conditions. This implies that T-duality should be realized also as a new symmetry principle in the effective field theory of D-branes in doubled space. That is, the effective double field theory for double D-branes should have T-duality as a manifest symmetry, and we expect it to simultaneously describe all D-brane configurations mutually related by T-duality transformations.
It is therefore of considerable interest to derive this theory from the doubled formalism, and we do so here, as well as discuss its properties. 

\subsection{Duality symmetric formulation and boundary gauge coupling}
\label{SecSymmetric}

The double field theory for the massless closed string sector was first discussed by Berman et al.~in \cite{Berman:2007xn,Berman:2007vi,Berman:2007yf} (see also \cite{Siegel:1993th,VanRaamsdonk:2003gj}). They considered a doubled formalism worldsheet theory analogous to the one described above in section~\ref{SecDGreview}, with target spacetime a doubled torus
fibred over some base manifold. In \cite{Berman:2007xn} it was shown that the vanishing one-loop beta function for the worldsheet doubled theory can be reduced by dimensional reduction to the usual background field equations for the standard sigma model.

On the other hand, the explicit form of a T-duality symmetric double field theory was later obtained by a different approach in a series of papers \cite{Hull:2009mi,Hull:2009zb,Hohm:2010jy,Hohm:2010pp}. They derived it by exploiting local gauge transformations and the symmetry algebra of double field theory in \cite{Kugo:1992md}. However, a strong constraint from the level-matching condition had to be imposed, and the target space interpretation of that constraint is obscure.  An attempt was made in \cite{Copland:2011yh} to show that the double field theories of \cite{Berman:2007xn} and \cite{Hohm:2010jy,Hohm:2010pp} are equivalent for special cases, but for the moment this would-be equivalence remains an interesting topic for further study.

    To derive a double field theory for our double D-branes, we should use one of the above approaches. However, it was shown in \cite{Coletti:2003ai} that it is nontrivial to derive the DBI action of D-branes from Witten's open string field theory \cite{Witten:1985cc}, and some numerical level truncation is necessary to determine the coefficients of higher derivative terms. It would therefore be difficult to generalize that derivation to the case of double field theory, unless the target space duality can be manifested in Witten's string field theory.  This is the reason we instead use the doubled formalism for the worldsheet theory to derive the effective theory for D-branes.
    
  We start by giving a brief review of the quantization of the worldsheet doubled formalism.
    The difficulty of quantizing the doubled  theory (\ref{doubledaction}) lies in how to incorporate the self-duality condition (\ref{selfdualcondition}) in a Lorentz invariant way. However, it turns out that the self-duality condition is equivalent to the chiral scalar conditions. To see this, consider the simple $2\times 2$-dimensional scenario described in section
\ref{SecExamples}, for which the self-duality condition reduces to the four equations (\ref{selfdualex1}). It is  easy to see that these conditions are equivalent to the following conditions for four chiral scalars $Z_{1\pm}$ and $Z_{2\pm}$,
\be\label{chiralselfdual}
\partial_{\pm}Z_{1\mp}=\partial_{\pm}Z_{2\mp}=0\;,
\ee
where
\be
Z_{1\pm}\equiv \tilde{X} \pm X-BY\;, \qquad Z_{2\pm}\equiv \tilde{Y}\pm Y + BX\;,
\ee
and $\partial_{\pm}\equiv \partial_0\pm \partial_1$. Eqn.~(\ref{chiralselfdual})  is the self-duality condition in the so-called ``chiral frame".
Note that also the  worldsheet action (\ref{doubledaction}) can be expressed in terms of chiral scalars; for the $2\times 2$-dimensional case we can define doubled coordinates $\mathbb{Z}\equiv (Z_{1+},Z_{2+},Z_{1-},Z_{2-})^t$ and rewrite the action in the chiral frame as
\be\label{chiralaction}
S_{w.s.}=\int d^2 \sigma \left(-\frac{1}{4} \eta^{\alpha\beta} \partial_{\alpha} \mathbb{Z}^t \partial_{\beta} \mathbb{Z}+ \frac{1}{2}\eta^{\alpha\beta}\partial_{\alpha}T\partial_{\beta}T \right) \;.
\ee

   In general, one can always go from the light-like frame, with $O(n,n)$-invariant metric (\ref{L}) and doubled space metric (\ref{Hmetric}), to the chiral frame where $\mathbb{L}$ and $\mathbb{H}$ are diagonal \cite{Berman:2007xn},
\be \label{chiralLH}
\mathbb{L}=\left( {\begin{array}{cc}
 \bid_{n \times n} & \bbz_{n \times n}  \\
 \bbz_{n \times n} & -\bid_{n \times n}  \\
 \end{array} } \right) \ , \hspace{1cm}
\mathbb{H} = \left(\begin{array}{cc}
\bid_{n \times n} & \bbz_{n \times n} \\
\bbz_{n \times n} & \bid_{n \times n} \\
\end{array}\right) \ .
\ee
Explicitly, we do this by use of vielbeins $\mathcal{V}_{\underline{I}}{}^I \in GL(2n)$, where $\underline{I},\underline{J}$ denote the chiral frame indices in doubled space; then the coordinates and the doubled metric transform as
\begin{equation}
\label{vielbein}
\begin{array}{rcl}
\mathbb{X}^{\underline{I}} &=& \mathcal{V}^{\underline{I}}{}_I \mathbb{X}^I\;,  \\
\mathbb{H}_{\underline{I}\underline{J}} &=& \mathcal{V}_{\underline{I}}{}^I \mathcal{V}_{\underline{J}}{}^J \mathbb{H}_{IJ} \ .
\end{array}
\end{equation}

It is clear from \eqref{chiralLH} that $(\mathbb{H}\pm\mathbb{L})$ behave as ``chiral projectors'',\footnote{See \cite{Jeon:2010rw,Jeon:2011kp,Jeon:2011cn} for the projection-compatible formalism of the double field theory.}
\be \label{prj}
  \frac{1}{2}\,(\mathbb{H}+\mathbb{L})
 = \left( {\begin{array}{cc}
  \bid_{n \times n} & \bbz_{n \times n}  \\
 \bbz_{n \times n} &  \bbz_{n \times n}  \\
 \end{array} } \right) \ , \hspace{1cm}
  \frac{1}{2}\,(\mathbb{H}-\mathbb{L})
 =\left( {\begin{array}{cc}
 \bbz_{n \times n} & \bbz_{n \times n}  \\
  \bbz_{n \times n} &  \bid_{n \times n}  \\
 \end{array} } \right)\,,
\ee
in the chiral frame. We will use this property to find the Green's
function in the chiral frame in Appendix~\ref{AppGreen}.

    Quantizing the worldsheet doubled theory is thus equivalent to quantizing the action of chiral scalars. By requiring that the generating functional in the doubled formalism be the same as that of the standard worldsheet formalism, the chiral scalar action can be written as a Floreanini-Jackiw (FJ)-type action \cite{Floreanini:1987as} as follows,
 %  \cite{Tseytlin:1990va,Berman:2007xn},
\be \label{FJ}
  S_{FJ} = \frac{1}{2} \int_{\Sigma} d^2\sigma [-\mathbb{H}_{IJ}\,\partial_1 \mathbb{X}^{I} \partial_1 \mathbb{X}^{J} + \mathbb{L}_{IJ}\; \partial_1 \mathbb{X}^{I} \partial_0 \mathbb{X}^{I} + \eta^{\alpha\beta}\partial_{\alpha}T \partial_{\beta}T] \ .
\ee
The action \eqref{FJ} is not manifest worldsheet Lorentz invariant.  This issue is well-known in quantization of self-dual theories, and has been extensively studied and finalized as the Pasti-Sorokin-Tonin (PST) formulation \cite{PST}.  Interestingly, even before the advent of the PST action, Tseytlin \cite{Tseytlin:1990va,Tseytlin:1990nb}  considered this problem and found that the condition of on-shell Lorentz invariance for the worldsheet theory (\ref{FJ}) requires that the symmetric matrices $\mathbb{H}$ and $\mathbb{L}$ satisfy
\be
\mathbb{L}=\mathbb{H} \; \mathbb{L}^{-1}\; \mathbb{H}\;.
\ee
This is just the statement that $\mathbb{H}$ must be a symmetric $O(n,n)$ matrix, whence follows that the action (\ref{FJ}) is $O(n,n)$ invariant.  By construction, the $\mathbb{H}$ and $\mathbb{L}$ given in (\ref{L}), (\ref{Hmetric}) and (\ref{chiralLH}) for both the light-like and chiral frames satisfy this Lorentz invariance condition. Hence the action (\ref{FJ}) is a consistent doubled formalism for quantization. Note that \eqref{FJ} is written in the light-like frame; if we work in the chiral frame, the indices in  \eqref{FJ} should be replaced with underlined ones.

 The consistency of the FJ-type action (\ref{FJ}) can also be justified by
using the PST formalism to quantize
the chiral fields. Taking the $n=2$ case chiral action \eqref{chiralaction} as an example, the PST formalism introduces auxiliary fields as Lagrange multipliers for the self-duality condition, or chiral conditions, in the original chiral action, 
\begin{eqnarray}
S_{PST}&=&\int d^2 \sigma \left(-\frac{1}{4} \eta^{\alpha\beta} \partial_{\alpha} \mathbb{Z}^t \partial_{\beta} \mathbb{Z}+ \frac{1}{2}\eta^{\alpha\beta}\partial_{\alpha}T\partial_{\beta}T \right)\nn \\
&-&\frac{1}{4} \int d^2 \sigma \;\sum_{i=1}^2 \left( \frac{\partial_+ a_{i+}}{\partial_- a_{i+}} (\partial_- Z_{i+})^2 +  \frac{\partial_- a_{i-}}{\partial_+ a_{i-}} (\partial_+ Z_{i-})^2 \right) \ ,\label{PSTaction}
\end{eqnarray}
where $a_{i\pm}$ are the Lagrange multipliers. By adding these nonlinear, Lorentz invariant gauge-fixing terms (second line of \eqref{PSTaction}) for the chiral condition to the action \eqref{chiralaction}, the PST action \eqref{PSTaction} acquires a  new gauge symmetry
\be
\delta a_{i\pm}=\Lambda_{i\pm}\;, \qquad \delta Z_{i,\pm} =\frac{\Lambda_{i\pm}}{\partial_{\mp} a_{i\pm} } \partial_{\mp}Z_{i\pm}\;, \qquad  i=1,2\;,
\ee
where $\Lambda_{i\pm}$ are the gauge parameters. This PST symmetry allows us to gauge away the non-chiral degrees of freedom. It is straightforward to generalize this procedure to cases with arbitrary $n$.  It is easy to see that the PST action \eqref{PSTaction} is manifestly Lorentz invariant.

We can now proceed in one of two ways: either covariantly quantize the theory by introducing the ghosts for the PST gauge symmetry, or gauge fix the PST action but break Lorentz invariance. Because our goal is to compute the one-loop beta function and not to obsess about covariance, we choose the latter approach. By choosing the Lorentz symmetry breaking condition $\partial_{\pm} a_{i\pm} = \partial_{\mp} a_{i\pm}$ to gauge fix the PST action \eqref{PSTaction}, the FJ-type action (\ref{FJ}) is obtained.

 Having thus arrived at a duality symmetric formulation of the quantum doubled theory, we need to add a boundary source term for the open string end-points to the FJ-type action. The action with this boundary term will then constitute our starting point for computing the effective double field theory for D-branes. First, however, we need to find the form of the boundary source term to add.

We propose the following boundary source term for the gauge fields,
\be \label{bdaction1}
 S_b = - \int_{\partial \Sigma} d\tau\: \left[ \mathbb{A}_I\, \partial_0 \mathbb{X}^I + A_T\;\partial_0 T \right]\;,
\ee
where the integral is carried out over the worldsheet boundary $\partial \Sigma$ parameterized by $\sigma^0\equiv\tau$. $\mathbb{A}_I$ is the doubled version of spatial boundary gauge field components,
\be
 \mathbb{A}_I \equiv  (A_i, \tilde{A}^i) \ ,
\ee
where the division of components is defined with respect to the polarization of $\mathbb{X}^I$ in \eqref{X}, and $A_T$ is the time-component of the gauge field.

  Since the end-points of the open string source the gauge fields on the worldvolume of the D-branes, only the Neumann part of the boundary gauge coupling is relevant for the derivation of the boundary effective theory. In fact, this statement follows automatically from the Dirichlet boundary condition \eqref{Dcond}. Inserting $\bid=\Pi_N^t + \Pi_D$ between $\mathbb{A}_I$ and $\partial_0 \mathbb{X}^I$ in \eqref{bdaction1}, the Dirichlet part vanishes due to the Dirichlet condition, while the Neumann part remains.\footnote{See eqn.~\eqref{326} for details.}

  Finally, before tackling the derivation  of a worldvolume effective double field theory for pairs of T-dual D-branes (double D-branes), we need to employ some further notation, for the convenience of expressing those formulae that are restricted to the Neumann subspace $\mathcal{N}$. We define the Neumann components of  $\mathbb{A}_I$ and $\mathbb{X}^I$ as $\mathcal{A}_p$ and $\mathcal{X}^p$, respectively,
  where the index $p$ labels the directions of the  $n$-dimensional spatial Neumann subspace $\mathcal{N}$, projected onto by $\Pi_N$. We thus have
\be
\mathbb{A}_I\, (\Pi_N^t)^I{}_J\, \partial_0 \mathbb{X}^J \equiv  \mathcal{A}_{p}\, \partial_0 \mathcal{X}^{p} \;.
\ee
We can thus write $\mathcal{A}\equiv \Pi_N \mathbb{A}$, and the
 derived field strength obeys the analogous projection condition $d\mathcal{A}\equiv \mathcal{F}=\Pi_N \mathbb{F} \Pi_N^t$, due to the assumption that $A_I$ depends only on the Neumann coordinates.\footnote{See section~\ref{SecDeriving} for  details.} Here $\mathbb{F}$ denotes the corresponding full doubled space field strength.

We can collect all the Neumann directions, including the time direction, in a new quantity,
\be\label{bfA}
\mathbf{A}_a\equiv (\mathcal{A}_p,A_T) \;, \qquad \mathbf{X}^a\equiv (\mathcal{X}^p,T)^t\;,
\ee
where the index $a$ labels both the temporal and spatial Neumann components. We moreover further decompose the spatial Neumann subspace $\mathcal{N}$, which is nothing but the spatial part of the double D-brane worldvolume,  into  two subspaces as
\be\label{NNbar}
\mathcal{N}=\mathcal{N}_s \oplus \tilde{\mathcal{N}_s}\;,
\ee
where the subspace $\mathcal{N}_s$ is parameterized by the Neumann subset of the physical coordinates \{$X^i$\}, and the subspace $\tilde{\mathcal{N}}_s$ is parameterized by the corresponding Neumann subset of \{$\tilde{X}_i$\}.

  Using this notation, the boundary action \eqref{bdaction1} can be rewritten as
\be\label{bdaction}
S_b= - \int_{\partial \Sigma} d\tau\: \left[ \mathcal{A}_{p}\, \partial_0 \mathcal{X}^{p} + A_T \partial_0 T + \ldots \right]
 = - \int_{\partial \Sigma} d\tau\: \mathbf{A}_{a}\, \partial_0 \mathbf{X}^{a} + \ldots
 \ ,
\ee 
where $\ldots$ denotes the (irrelevant) Dirichlet part. This looks very much like the conventional worldsheet formalism; recall, however, that the index $p$ (or $a$) labels the Neumann components of the pair of T-dual  D-branes, and is a doubled index. That is, $\mathbf{A}(\mathbf{X})$ is the double gauge field on the  worldvolume of double D-branes.

\subsection{Deriving the effective double field theory}
\label{SecDeriving}

We are now ready to derive the effective double field theory based on the actions (\ref{FJ}) and (\ref{bdaction}), by evaluating the one-loop beta function of the boundary gauge coupling. For this purpose we generalize the background field method of \cite{Abouelsaood:1986gd,Mukhi:1985vy} to doubled formalism.

  Consider some quantum fluctuations $\xi$ over the classical background fields $\mathbb{X}$ and $T$,
\be \label{bfe}
\mathbb{X}^I \to \mathbb{X}^I + \xi^I \ , \hspace{1cm}
T \to T + \xi^T \ ,
\ee
where $\xi^I$ have doubled degrees of freedom while $\xi^T$ is the ordinary non-doubled fluctuation. We assume a flat target space; see \cite{Berman:2007xn, Mukhi:1985vy} for the background field expansion in a generic curved target space. Moreover, for convenience we work in Euclidean worldsheet signature,\footnote{The Wick rotation on the worldsheet time is $\tau_M \to - \mathrm{i}\tau_E$, so that the action transforms as $\mathrm{i}\,S_M = -S_E$, where the subscripts $M$ and $E$ denote Minkowskian and Euclidean, respectively. In the main text we omit the subscripts.} and we now let $\sigma^0\equiv\tau$ and $\sigma^1\equiv\sigma$ denote the Euclidean string worldsheet coordinates. With the substitution (\ref{bfe}), the bulk action (\ref{FJ}) expands to
\bea \label{bfeBulk}
 S_E &=& \frac{1}{2} \int_{\Sigma} d^2 \sigma \; \left\{(\mathbb{H}_{IJ}\,\partial_1 \mathbb{X}^I \partial_1 \mathbb{X}^J
  -\mathrm{i}\, \mathbb{L}_{IJ}\; \partial_1 \mathbb{X}^I \partial_0 \mathbb{X}^J - \delta^{\alpha\beta} \partial_{\alpha} T \partial_{\beta} T) \right.\nonumber\\
 && \hspace{1cm} + \;( 2\mathbb{H}_{IJ}\,\partial_1 \xi^I \partial_1 \mathbb{X}^J
       -\mathrm{i} \mathbb{L}_{IJ}\; \partial_1 \xi^I \partial_0 \mathbb{X}^J -\mathrm{i} \mathbb{L}_{IJ}\; \partial_1 \mathbb{X}^I \partial_0 \xi^J - 2 \delta^{\alpha\beta} \partial_{\alpha} T \partial_{\beta} \xi^T  )\\
&& \hspace{2cm} \left. +( \mathbb{H}_{IJ}\,\partial_1 \xi^I \partial_1 \xi^J -\mathrm{i} \mathbb{L}_{IJ}\; \partial_1 \xi^I \partial_0 \xi^J - \delta^{\alpha\beta} \partial_{\alpha}\xi^T \partial_{\beta} \xi^T )+{\cal O}(\xi^3)\nonumber
\right\} \ .
\eea
  Similarly expanding the boundary action (\ref{bdaction})  yields
\bea \label{bfeBdy}
S_{Eb}&=&\mathrm{i}\,\int_{\partial \Sigma} d\tau \; \Big\{ \mathbf{A}_a \, \partial_0 \mathbf{X}^a + (\xi^a\, \mathbf{F}_{ab} \, \partial_0 \mathbf{X}^b )+\frac{1}{2} \left(\xi^c \xi^a \, \nabla_c \mathbf{F}_{ab}
   \partial_0 \mathbf{X}^b + \xi^a \partial_0 \xi^b \mathbf{F}_{ab} \right) \nn\\
&&\hspace{1.8cm} +\frac{1}{3} \left(\frac{1}{2}\xi^c \xi^d \xi^a \nabla_c \nabla_d \mathbf{F}_{ab} \partial_0 \mathbf{X}^b + \xi^c \xi^a \partial_0 \xi^b \nabla_c \mathbf{F}_{ab}\right) +\; {\cal O}(\xi^4) + \ldots \Big\}\ ,
\eea
where $\ldots$ represents the (irrelevant) Dirichlet part. Here we have introduced the ``Neumann notation'', described in section~\ref{SecSymmetric}, for the Neumann components of the fluctuation fields, including both temporal and spatial ones, thus: $\xi^a=(\xi^p,\xi^T)^t$.  The expansions and contractions above are done entirely in the Neumann subspace, except the $\ldots$ part. In the following we further adopt the slowly-varying field approximation, so that the contribution from the $\nabla \mathbf{F}$ terms is less significant  than that of the $\mathbf{F}$ terms, and may be treated as the interaction; terms with higher order derivatives of $\mathbf{F}$ can be neglected.

  Note that we assume the gauge field is  a function of the Neumann coordinates only, $\mathbf{A}_a=\mathbf{A}_a(\mathbf{X}^a)$, or in other words, $\mathbf{A}=\Pi_N \mathbb{A}(\Pi_N^t \mathbb{X},T)$ and $A_T=A_T(\Pi_N^t \mathbb{X},T)$.  This is because the gauge field is living on the worldvolume of the double D-branes so that it only depends on the Neumann coordinates. With this assumption, we still maintain the T-dual covariance on the  Neumann subspace $\mathcal{N}=\mathcal{N}_s \oplus \tilde{\mathcal{N}_s}$, and we need not further assume that $\mathbf{F}_{ab}=0$ for $a \in \mathcal{N}_s$, $b \in \tilde{\mathcal{N}_s}$. Therefore, the resultant effective double field theory should be both gauge and T-duality covariant.

  From the first order terms in the expansions above we find the bulk equations of motion for the background fields $\mathbb{X}^J$ and $T$,
\bea
\partial_1 (\mathbb{H}_{IJ}\, \partial_1\mathbb{X}^J ) -\mathrm{i}\, \mathbb{L}_{IJ} \partial_0 \partial_1 \mathbb{X}^{J}&=& 0 \ , \label{eom} \\
\delta^{\alpha\beta} \partial_{\alpha}\partial_{\beta} T &=& 0 \ . \label{eomT}
\eea
The corresponding Neumann boundary conditions\footnote{We choose $T$ to satisfy the Neumann boundary conditions, since this is the case for all D-branes but the D-instanton.} read
\bea
\mathcal{H}_{pq} \partial_1 \mathcal{X}^q + \mathrm{i}\,\mathbf{F}_{pa} \partial_{0} \mathbf{X}^a \;|_{\partial \Sigma} \!\!\! &=& \!\!\! 0 \hspace{0.2cm}\ , \label{bcn} \\
- \partial_1 T + \mathrm{i}\,\mathbf{F}_{Ta} \partial_{0} \mathbf{X}^a \; |_{\partial \Sigma}  \!\!\!&=&\!\!\! 0 \hspace{0.2cm}\ , \label{bcnT}
\eea
and the Dirichlet boundary conditions are
\be\label{DconT}
\Pi_D\; \xi \; |_{\partial \Sigma}=0\;, \qquad \Pi_D\; \mathbb{X} \; |_{\partial \Sigma}=0\;.
\ee
Here we have introduced yet another quantity, the pull-back  $\mathcal{H}\equiv \Pi_N \; \mathbb{H}\; \Pi^t_N$  of the doubled space metric $\mathbb{H}$ to the Neumann subspace $\mathcal{N}$.   Note also that the pull-back of the $O(n,n)$ metric $\mathbb{L}$ to the Neumann subspace is zero by the null condition (\ref{nullc}). To arrive at the first term on the left-hand side of (\ref{bcn}) we have used the orthogonality condition (\ref{orthocon}) and Dirichlet boundary condition (\ref{DconT}) to write
\be \label{326}
\Pi_N \; \mathbb{H}\; \partial_1 \mathbb{X}=\Pi_N \; \mathbb{H} \; (\Pi_N^t+\Pi_D) \; \partial_1 \mathbb{X}=\mathcal{H}\; \partial_1 \mathcal{X}\;,
\ee
and to cancel the $\mathbb{L}$ term on the boundary.

  Similarly, the second order terms in the expansions \eqref{bfeBulk} and \eqref{bfeBdy} provide the equations of motion and boundary conditions for the fluctuation fields $\xi$, and their form is the same as that of the background fields.  Based on these equations of motion and boundary conditions we can write down the Green's equations and their boundary conditions. However, the equations of motion can be put into a unified form if we introduce  \emph{extended} metrics \cite{Berman:2007xn},
 \be \label{Oin1G}
 \hat{\mathbb{H}}_{A B} \equiv
 \left( {\begin{array}{cc}
 \mathbb{H}_{IJ} & 0  \\
 0 & g_{\scriptscriptstyle{TT}}  \\
 \end{array} } \right) \;, \hspace{1cm}
\hat{\mathbb{L}}_{AB} \equiv
 \left( {\begin{array}{cc}
 \mathbb{L}_{IJ} & 0  \\
 0 & 0  \\
 \end{array} } \right)\;,
\ee
where the time-time component $g_{\scriptscriptstyle{TT}}$ of the doubled metric is usually set to $-1$ for Minkowskian target spacetime. These are metrics with respect to the extended coordinates $\hat{\mathbb{X}}^A \equiv (\mathbb{X}^I,T)^t$.   We can moreover define the extended Neumann projector
\be\label{largePN}
(\hat{\Pi}_N)_{A}{}^{B}\equiv \left( {\begin{array}{cc}
 (\Pi_N)_{I}{}^J & 0  \\
 0 & 1  \\
 \end{array} } \right) \;,\ee
which by definition satisfies $\hat{\Pi}_N^2=\hat{\Pi}_N$.  Using the general form of $\Pi_N$ given in \eqref{PNchiral} and $\mathbb{H}$ given in \eqref{chiralLH}, one can furthermore show that from $\Pi_N\; \mathbb{H}\; \Pi_N^t=\Pi_N\; \mathbb{H}$ (implied by orthogonality) follows that
\be\label{PNp}
\hat{\Pi}_N\; \hat{\mathbb{H}}\; \hat{\Pi}_N^t=\hat{\Pi}_N \; \hat{\mathbb{H}} \;.
\ee

The boundary conditions can also be put into a unified form if we introduce the extended pull-back doubled metric to the Neumann subspace,
\be \label{Oin2G}
 \mathbf{g}_{ab} \equiv
 \left( {\begin{array}{cc}
 \mathcal{H}_{pq} & 0  \\
 0 & g_{\scriptscriptstyle{TT}}  \\
 \end{array} } \right) \;, \qquad \mbox{or formally}  \quad \mathbf{g}=\hat{\Pi}_N \; \hat{\mathbb{H}} \; \hat{\Pi}_N^t\;.
\ee
We can moreover pull back the field strength and define its extended partner as
\be
\hat{\mathbb{F}}_{AB}\equiv
 \left( {\begin{array}{cc}
 \mathbb{F}_{IJ} & F_{IT}  \\
F_{TJ} & F_{\scriptscriptstyle{TT}}  \\
 \end{array} } \right) \qquad \mbox{so that} \quad  \mathbf{F}=\hat{\Pi}_N \; \hat{\mathbb{F}}\; \hat{\Pi}_N^t\;.
\ee
From here on we will use the metric $\mathbf{g}_{ab}$ as the metric on the Neumann subspace, i.e., on the worldvolume of the double D-branes. That is, we will use it to raise and lower the Neumann indices. We shall therefore require that we can write $\mathbf{g}^{ab}=(\mathbf{g}^{-1})^{ab}$. At  first sight this seems not to be the case, since by definition
\be
\mathbf{g}^{-1}\equiv \Pi_N^t \; \mathbb{L}^{-1} \mathbb{H} \; \mathbb{L}^{-1} \; \Pi_N =\Pi_N^t \; \mathbb{L}^{-1}\;  \Pi_D^t \;\mathbb{H} \; \Pi_D  \; \mathbb{L}^{-1} \; \Pi_N \;,
\ee
which involves the Dirichlet sector of the doubled metric and yields
\be
\mathbf{g}\; \mathbf{g}^{-1}=\Pi_N \neq \bid \;,
\ee
\be
\mathbf{g}^{-1}\; \mathbf{g}=\Pi_N^t \neq \bid \;.
\ee
However, because $(\Pi_N)_{A}{}^{B} = (\Pi_N)_{A}{}^{C} \delta_C{}^D (\Pi_N)_{D}{}^{B}$ and $(\Pi_N^t)^{B}{}_{A} =
(\Pi_N^t)^{B}{}_{C} \delta^C{}_D (\Pi_N^t)^{D}{}_{A}$, we can treat $\Pi_N$ and $\Pi_N^t$ as
the identity whenever the consideration is restricted to the Neumann subspace. And since in the Neumann Green's function analysis we are concerned only with the Neumann subspace, we can thus freely use $\mathbf{g}^{ab}$ as a metric inverse in our calculations.

In the extended notation, the doubled space Green's function $G^{AB}$  satisfies the unified equations of motion,
\be \label{Oin1eom}
\left(\, \delta_A{}^T \delta_B{}^T \, \hat{\mathbb{H}}_{TT} \partial_{0}^{2} + \hat{\mathbb{H}}_{AB} \partial_1^2  - \mathrm{i} \hat{\mathbb{L}}_{AB} \partial_0 \partial_1 \, \right) G^{BD}(\vec{\sigma},\vec{\sigma}') = -
\delta_A{}^D \, 2\pi\,\delta(\vec{\sigma}-\vec{\sigma}') \ ,
\ee
and the unified Neumann boundary conditions,
\be\label{Unify N}
 \mathbf{g}_{ab} \; \partial_1 G^{bc}(\tau,\tau') + \mathrm{i} \mathbf{F}_{ab}\; \partial_0 G^{bc}(\tau,\tau')\Big|_{\sigma = 0} = 0 \ .
\ee

   After solving for the Green's function, we can then use it to evaluate the one-loop counter term for the boundary gauge coupling, which takes the form
\be\label{counterterm}
\frac{i}{2} \int_{\partial \Sigma} d\tau\; \mathbf{\Gamma}_a \partial_0 \mathbf{X}^a \;.
\ee
The relevant leading contribution to the one-loop counter term comes from the second order interaction in the expanded boundary action (\ref{bfeBdy}),
\be\label{3rdint}
 S_{\mathrm{int}} = \frac{\mathrm{i}}{2} \int_{\partial \Sigma} d\tau\; \xi^c \xi^a \; \nabla_c \mathbf{F}_{ab}\; \partial_0 \mathbf{X}^b\;.
\ee
By comparing the forms of (\ref{counterterm}) and (\ref{3rdint}), or by reading from the one-loop diagram in Fig.~2 of \cite{Abouelsaood:1986gd}, we find that the counter term is given by
\be\label{countertermG}
\mathbf{\Gamma}_a\equiv  \lim_{\epsilon \rightarrow 0} \; G^{bc}(\epsilon\equiv \tau-\tau') \nabla_b \mathbf{F}_{ca}\;,
\ee
where $\epsilon$ is the short-distance UV cutoff.
  Then the beta function for the boundary gauge coupling may be obtained as
\be
\mathbf{\beta}_a\equiv -2\pi \epsilon \frac{\partial \mathbf{\Gamma}_a}{\partial \epsilon}\;.
\ee
If we demand Weyl invariance on the worldsheet boundary, the equations of motion for the effective double field theory of D-branes are just
 \be
 \mathbf{\beta}_a=0\;.
 \ee
 
 Note that, although our expanded boundary action is formally the same as the one in the standard worldsheet formalism, the worldsheet action (which takes the FJ form) is different from the usual one. Therefore, we  expect to obtain an effective field theory  different from the usual DBI action.

   Next we solve (\ref{Oin1eom}) and (\ref{Unify N}) for the Green's function. Because the doubled formalism is $O(n,n)$ covariant, we can choose a particular $O(n,n)$ frame in which to solve for the Green's equation. Since in the chiral frame the metrics $\mathbb{H}$ and $\mathbb{L}$ are diagonal, solving the Green's equation in this frame should be quite straightforward.  The transformation from the light-like frame to the chiral frame is achieved by use of the vielbeins given in (\ref{vielbein}). These vielbeins, which also ``rotate'' $\xi^I$, involve the $B$-field components which are constant in our setup. Unlike in \cite{Berman:2007xn}, however, the vielbeins here do not contribute to the dynamics. 
   
   Thus our approach is to solve the Green's function in the chiral frame and then transform it back to $O(n,n)$ covariant form in the light-like frame. 
   We defer the explicit calculation to Appendix~\ref{AppGreen}.
   Up to a term $\mathbf{u}^{ab}$, which will be determined later by the boundary conditions, the Neumann Green's function solution is obtained as
\be \label{Gansatz}
 G^{ab} = \mathbf{g}^{ab} \, \Delta_0  +\mathbf{u}^{ab}
\ee
where
\be
 \Delta_0 \ \equiv \  -\frac{1}{4\pi}[\ln(z -z')+ \ln (\bar{z}-\bar{z}') ] \ .
\ee
Here we have introduced the complexified worldsheet coordinates   $z \equiv \sigma + \mathrm{i}\tau$, $\bar{z}\equiv\sigma - \mathrm{i}\tau$, so as to represent the worldsheet as the right half of the $z$-plane, with the boundary defined by the imaginary axis.
In terms of $z$ and $\bar{z}$, the Neumann boundary condition becomes
\be \label{Oin1bc}
 (\mathbf{g}- \mathbf{F})_{ab} \,\partial_z G^{bc}(z,z')
 + (\mathbf{g}+ \mathbf{F})_{ab}\, \partial_{\bar{z}} G^{bc}(z,z') \; |_{z=-\bar{z}}= 0 \ ,
\ee
from which we can determine the mirror charge part $\mathbf{u}^{ab}$ of the Green's function. The result is
\be\label{uab}
 \mathbf{u}^{ab} =
 -\frac{1}{4\pi} \left\{  \left(\frac{\mathbf{g}+\mathbf{F}}{\mathbf{g}-\mathbf{F}}\right)^a{}_c \, \mathbf{g}^{cb} \;\, \ln (z+\bar{z}') + \left(\frac{\mathbf{g}-\mathbf{F}}{\mathbf{g}+\mathbf{F}}\right)^a{}_c \, \mathbf{g}^{cb} \, \ln (\bar{z}+z') \right\}\;.
\ee
In this expression  we have employed the abbreviation
$\frac{\mathbf{C}}{\mathbf{D}} \equiv  \mathbf{D}^{-1} \mathbf{C}$, and all the inverses are taken within the Neumann subspace.

    If we substitute (\ref{uab}) in (\ref{Gansatz}), we can identify the counter term defined in (\ref{countertermG}). We find
\be\label{finalcounterterm}
\mathbf{\Gamma}_a
= -\frac{1}{2\pi}\left\{ \mathbf{g}^{-1} +\frac{1}{2} \left( \frac{\mathbf{g}+\mathbf{F}}{\mathbf{g}-\mathbf{F}}\right)\,\mathbf{g}^{-1} +\frac{1}{2} \left( \frac{\mathbf{g}-\mathbf{F}}{\mathbf{g}+\mathbf{F}}\right)\,\mathbf{g}^{-1}
\right\}^{bc} (\nabla_b \mathbf{F}_{ca} ) \ln \epsilon + \ldots \ ,
\ee
where $\epsilon$ is the short distance UV cutoff, and $\ldots$ represents the finite part in the limit $\epsilon \to 0$. As a result, the equations of motion of the effective double field theory for double D-branes read
\be \label{effeom}
\beta_a\equiv \left\{  \left( \mathbf{g}-\mathbf{F}^2\right)^{-1} \right\}^{bc} \nabla_b \mathbf{F}_{ca} =0\;.
\ee
This is exactly the equation found by Abouelsaoodas et al.~in \cite{Abouelsaood:1986gd}, from the standard worldsheet formalism.

   By the same arguments as those of \cite{Abouelsaood:1986gd}, the equations of motion on the form
 \be
 \left\{  \left( \mathbf{g}-\mathbf{F}^2\right)^{-1} \right\}^{ab}\beta_b=0
\ee
 can be obtained from a DBI-like action
\be \label{Seff}
S_{eff}=\int d^{n+1}\mathbf{X}\; \sqrt{-\det(\mathbf{g}+\mathbf{F})}  \;,
\ee
where the determinant is defined within the Neumann subspace. This is the effective double field theory for the double D-branes. In the above derivation, we have assumed that the extended double field strength satisfies the Bianchi identity, i.e.,
\be
\nabla^a \mathbf{F}^{bc}+\nabla^b \mathbf{F}^{ca}+\nabla^c \mathbf{F}^{ab}=0\;.
\ee

\subsection{Comments on the effective double field theory}

  Although the form of (\ref{Seff}) naively looks like that of the DBI action for ordinary D-branes, there are important differences, which we elaborate on in this section. 

\bigskip

  {\bf 1. Gauge symmetry:} Because the effective action is of the DBI form, the theory is invariant under the extended doubled space gauge transformation
\be\label{gaugeTN}
\mathbf{A}_a \longrightarrow \mathbf{A}_a + \nabla_a \mathbf{\Lambda} \,,
\ee
where the gauge parameter $\mathbf{\Lambda}$ is a function of the double coordinates on the Neumann subspace, i.e.,
$\mathbf{\Lambda}=\mathbf{\Lambda}(\mathbf{X^a})$.  This is the generalization of the usual gauge symmetry.

\bigskip

  {\bf 2. $O(n,n)$ symmetry:} Although the action (\ref{Seff}) is T-duality invariant on the worldvolume of the double D-branes, it is not manifestly $O(n,n)$ invariant. However, we can make the $O(n,n)$ symmetry manifest by recovering the Neumann and Dirichlet projectors explicitly. Moreover, by using $O(n,n)$ covariance and the property (\ref{PNp}), the extended pull-back metric can be further reduced to
\be
\mathbf{g}=\hat{\Pi}_N \mathbb{H}\;.
\ee
The general form of the projector $\hat{\Pi}_N$ is worked out in Appendix \ref{AppOnn} in both the chiral and  light-like frame. Then the manifest $O(n,n)$ symmetric action for the double D-branes can be written as
\be\label{SeffOnn}
S_{eff}= \int  \frac{ d^{n+1} (\hat{\Pi}_N \hat{\mathbb{X}})}{Vol^n[\Pi_D]} \; \sqrt{-\det( \Pi_D+ \hat{\Pi}_N\; \hat{\mathbb{H}} + \hat{\Pi}_N \; \hat{\mathbb{F}} \; \hat{\Pi}_N^t )} \,,
\ee
where  $Vol^n[\Pi_D]$ denotes the $O(n,n)$ covariant volume   of  the space of Dirichlet projectors. The Dirichlet projector $\Pi_D$ appears because computing the determinant requires full rank matrices, but its contribution is canceled out by the factor $1/Vol^n[\Pi_D]$. The gauge transformation (\ref{gaugeTN}) can similarly be written on an $O(n,n)$ covariant form.
Because of its $O(n,n)$ covariant form, the effective action (\ref{SeffOnn}) may be described as a master action for all  D-brane configurations related by T-duality transformations.

\bigskip

  {\bf 3. Reduction by eliminating the T-dual components:}   The doubled formalism is classically equivalent to the conventional worldsheet formalism \cite{Hull:2004in, Hull:2006va, Thompson:2010sr}.  Although there appear to be double degrees of freedom in the doubled formalism, half of them are eliminated by imposing the self-duality condition. More precisely, we can use the self-duality constraint to eliminate the dual coordinates $\tilde{X}_i$ in favor of the coordinates $X^i$. We expect the equivalence to hold also at the quantum level, and this is indeed the case for  closed string theory \cite{Copland:2011yh,Berman:2007vi}. We may ask if we can also reduce our effective doubled field theory (\ref{Seff}) to the conventional D-brane DBI action by applying the self-duality condition.

   In terms of the Neumann and Dirichlet subspaces in our formalism, the self-duality condition is just the interchange of  Neumann  and Dirichlet coordinates. This can be seen by using the null condition (\ref{nullc}) and the orthogonality condition (\ref{orthocon}) to rewrite the self-duality condition (\ref{selfdualcondition}) on the worldsheet boundary as
\be
\Pi^t_N\; \partial_0 \mathbb{X} |_{\partial \Sigma} = \Pi_N^t \; \mathbb{L}^{-1} \; \Pi_D^t \; \mathbb{H}\; \Pi_D\; \partial_1 \mathbb{X} |_{\partial \Sigma}\;.
\ee
However, because our boundary action does not sustain the dynamics along the Dirichlet directions, the self-duality condition simply kills the gauge dynamics along the Neumann directions, i.e., the gauge potential is no longer a function of the Neumann coordinates.
Hence, if we want to reproduce the conventional D-brane theory, we can  require that the gauge fields  depend on  either $\mathbf{X}_a \in \mathcal{N}_s$ or  $\mathbf{X}_a \in \tilde{\mathcal{N}}_s$, but not on both.  This is analogous to the strong constraint derived from the closed field theory as discussed in \cite{Hull:2009mi} to obtain the effective double field theory for gravity. However, in our case it is inferred from the self-duality condition.

This property of the gauge fields allows us, if there is no electric field, i.e., if $F_{pT}=0$,  to reduce the effective action (\ref{Seff}) to the conventional DBI action for a single brane in the $B=0$ case.  This can be straightforwardly verified since our effective action has a DBI form up to some proper block diagonalization of $\mathbf{g}$ and $\mathbf{F}$ (a consequence of the need to calculate the determinant in the Neumann subspace).\footnote{In Appendix \ref{AppO22} we see that there are some examples for which $\mathbf{g}$ and $\mathbf{F}$ are not on a block-diagonal form.}

This statement can be illustrated in an elucidating manner by the example of a double D2-brane in $2\times 2$ doubled space, with its worldvolume along the $(X,Y)$ directions, coupled to a constant background $B$-field. The corresponding Neumann and Dirichlet projectors are given by the Type II solution in Appendix \ref{AppO22}. After a proper Jordan decomposition to express the Neumann subspace in the block-diagonal form of $\mathbf{g}+\mathbf{F}$, the square of the DBI-like double field action is given by
\be
 \det (\mathbf{g}+\mathbf{F}) =
 -\frac{(1 + 3 B^2 + B^4)^2}{(1 + B^2)^4}\, \{1 + 2 B^2 + B^4 - (1+B^2)(F_{XT}^2 + F_{YT}^2 )+ F_{XY}^2 \} \ .
\ee
Here the determinant is taken within the Neumann subspace. On the other hand, the conventional DBI determinant associated with the D2-brane configuration at hand reads
\be
-1 - B^2 + F_{XT}^2 + F_{YT}^2 - 2 B F_{XY} - F_{XY}^2 \ .
\ee
Thus we see that our double effective DBI-like action cannot be reduced to the conventional DBI action when $B \neq 0$.
However, when $B=0$ we have
\be
 \det (\mathbf{g}+\mathbf{F}) =
 -1 + F_{XT}^2  + F_{YT}^2 - F_{XY}^2 \ ,
\ee
which is identical to the conventional DBI determinant for $B=0$.

Besides the reason mentioned above (coordinate dependence of the gauge fields), our inability to recover the standard DBI action when $B\neq 0$ may also be due to a lack of compensation between the worldsheet and the boundary. 
Because we do not double the time coordinates, there is no explicit T-dual covariance associated with the time component. This causes nontrivial mixing of the electric fields associated with different D-branes in the effective action, i.e., there are terms like $F_{Ti}F_{T\tilde{j}}$, which prevent a reduction to the single brane action. To avoid this obstacle, we need to restrict the electric field to live only on the D-brane worldvolume $\in \mathcal{N}_s$, to which we want to reduce the double action, and not on the T-dual counterpart $\in \tilde{\mathcal{N}}_s$. In this way we can perform the reduction, and the resulting action will be a standard DBI action for the conventional D-brane.

\bigskip

  {\bf 4. $B$-field dependence and bulk gauge symmetry:} The dependence on the constant $B$-field in the effective action is encoded in the projector $\hat{\Pi}_N$, and this dependence cannot be of the same simple form $\sqrt{-\det (g+B+F)}$ as in the conventional theory for a single D-brane.  Because the $B$-field dependence involves the worldsheet gauge symmetry as discussed in \cite{Hull:2009mi,Hull:2009zb}, it is interesting to see how the boundary gauge symmetry can be related to the worldsheet one.

  We sketch this  issue as follows. Hull and Zwiebach constructed an $O(n,n)$-covariant action for closed string field theory based on the doubled metric $\mathcal{H}$ which is an $O(n,n)$-tensor \cite{Hohm:2010pp}. This double action is gauge invariant under a local diffeomorphism $\delta_{\xi}$ which realizes a generalized Lie derivative on the doubled geometry, defined by
\begin{equation} \label{Lieg}
\delta_{\xi} \mathcal{H}_{MN} =\hat{\mathcal{L}}_{\xi}\mathcal{H}_{MN}\equiv\xi^{P}\partial_{P}\mathcal{H}_{MN}+(\partial_{M}\xi^{P}-\partial^{P}\xi_{M})\mathcal{H}_{PN}+(\partial_{N}\xi^{P}-\partial^{P}\xi_{N})\mathcal{H}_{MP}.
\end{equation}
$\hat{\mathcal{L}}$ satisfies
\begin{equation}
[\hat{\mathcal{L}}_{\xi_{1}},\hat{\mathcal{L}}_{\xi_{2}}]=-\hat{\mathcal{L}}_{[\xi_{1},\xi_{2}]_{C}},
\end{equation}
where $[ \ , \ ]_{C}$ is called a $C$-bracket and was first introduced by Siegel \cite{Siegel:1993th}. It is defined as
\begin{equation}
([\xi_{1},\xi_{2}]_{C})^{M}=[\xi_{1},\xi_{2}]-\frac{1}{2}(\xi_{1}^{N}\partial^{M}\xi_{2N}-\xi_{2}^{N}\partial^{M}\xi_{1N}).
\end{equation}
Note that the gauge invariance of the action and the closure of the gauge algebra require that arbitrary fields $A$ and $B$ satisfy $\partial^{M}\partial_{M}A=0$ and $\partial^{M}A\partial_{M}B=0$. These constraints arise from the level matching condition in the closed string theory. 

It was observed in \cite{Hull:2009zb} that the $C$-bracket on doubled geometry is equivalent to the Courant bracket on generalized geometry. Like the C-bracket, the Courant bracket does not satisfy the Jacobi-identity, but instead some specific Jacobiator relations.

In our formulation of the open string worldsheet action on doubled geometry, we expect the same underlying mathematical structure, i.e., we expect the action to be invariant under a local diffeomorphism and that the gauge algebra is closed under the $C$-bracket. The worldsheet action (\ref{doubledaction}) together with the boundary action (\ref{bdaction1}) and the self-duality condition  (\ref{selfdualcondition}) should manifestly exhibit both $O(n,n)$ covariance and gauge symmetry, and from this action we would expect to be able to derive the relation between the worldsheet and the boundary gauge symmetry. However, straightforward verification shows that neither the worldsheet action nor the self-duality condition are invariant under the gauge transformation (\ref{Lieg}). It remains an open issue to find the worldsheet gauge symmetry for the doubled formalism of the worldsheet description.

\section{Conclusions}
\label{SecConclusion}

We considered the doubled formalism for open strings on D-branes in flat space with a constant $B$-field and derived the  effective double field theory for this configuration. 
Doubled formalism for open strings has been studied in the past, but then mainly focusing on the kinematical constraints on the embedding of D-branes in doubled geometry. What is new about our study is that, while clarifying the role of doubled formalism as a unifying description of pairs of T-dual D-branes (double D-branes), we explored the dynamical aspects of the theory.

  First, we explicitly demonstrated the unifying power of the doubled formalism in describing double D-branes via boundary conditions, by exploiting the self-duality constraint to change viewpoints. We then constructed the general form of the Neumann projectors and used them to formulate the boundary action for open strings on double D-branes. Finally, we applied the background field method to the open string doubled formalism to derive the effective double field theory for double D-branes. It turns out that the effective action takes the DBI form in the appropriate notation for the worldvolume metric and field strength. A possible direction for further study would be to evaluate the higher order correction to this DBI action, by considering the contribution to the beta function of the Neumann Green's function on the annulus.

  Although our effective action takes a simple DBI form, it is $O(n,n)$ and gauge invariant, and it is a double action. We hope this new action will inspire new progress in relation to  the dynamical aspects of  manifest T-duality covariance. Several important topics, such as non-commutative doubled geometry and tachyon condensation of double D-branes, should be investigated further in the context of doubled formalism.  Moreover, as discussed in the previous section, the issue concerning $C$-bracket gauge invariance of our double theory remains open. This issue is related to the explicit dependence of the $B$-field in the effective action. In conclusion, we hope our results will constitute a basis  for future efforts toward an $O(n,n)$ perspective of string theory.

\section*{Acknowledgments}
C.A.~is grateful to her employer Lunascape Corporation for their kind permission to use their address as affiliation. However, the work of C.A.~has been conducted entirely on her own time and equipment; no part of this research has been in any way supported by Lunascape Corporation, through financial or other means. S.H.D.~wishes to thank Shoichi Kawamoto for  helpful discussions. P.K.~would like to thank Ryszard Nest for useful conversations. The work of P.K.~is supported by the JSPS fellowship.  F.L.L.~wishes to thank for the warm hospitality of ICTS at USTC, where he gave a talk on this paper. F.L.L.~is also grateful to the College of Physics of Peking University for their hospitality and kind support. The work of S.H.D.~and F.L.L.~is supported by Taiwan's NSC Grant No.~99-2811-M-003-005 and No.~97-2112-M-003-003-MY3.  We are also grateful for the support of NCTS.

\appendix

\section{Derivations from the chiral frame}
\label{AppOnn}

 In this section, we derive the most general form that a Neumann projector is allowed in the chiral frame, and then we transform this solution to the original light-like frame. Because the doubled metric $\mathbb{H}$ and the $O(n,n)$ invariant metric $\mathbb{L}$ are diagonal in the chiral frame where the $B$-field vanishes, some computations are less cumbersome to do in the chiral frame, and the underlying physics for our double field theory becomes easier to interpret. We set the target space metric flat as usual.

We introduce a vielbein set $\mathcal{V}^{\underline{I}}{}_{I}$ to transform between the light-like $O(n,n)$ representation (with indices $I,J$) and the chiral frame one (with indices $\underline{I}, \underline{J}$), so that the doubled metric $\mathbb{H}$ (\ref{Hmetric}) can be decomposed as \cite{Hull:2004in}
\be
 \mathbb{H} = \mathcal{V}^t \mathcal{V} \ .
\ee
Explicitly, the vielbein components may be written as
\be \label{vb}
\mathcal{V}^{\underline{I}}{}_{I} = \frac{1}{\sqrt{2}}
\left( \begin{array}{cc}
  \delta^i{}_j - B^i{}_j &  \delta^{ij}  \\
  - \delta_{ij} - B_{ij}           &  \delta_i{}^j
\end{array}\right) \ , \qquad\quad
\mathcal{V}^I{}_{\underline{I}} = \frac{1}{\sqrt{2}}
\left( \begin{array}{cc}
  \delta^i{}_j &  -\delta^{ij}  \\
  \delta_{ij} + B_{ij}   &  \delta_i{}^j - B_i{}^j
\end{array}\right) \ ,
\ee
where $B_{ij}$ denote the components of the $n\times n$ $B$-field matrix, and we have normalized the vielbein by setting $\det \mathcal{V} =1$.
Given this set of vielbeins, the form (\ref{chiralLH}) of $\mathbb{H}_{\underline{I}\underline{J}}$ and $\mathbb{L}_{\underline{I}\underline{J}}$ in the chiral frame follows straightforwardly.

The general form of the Neumann projector in the chiral frame can be expressed as
\be \label{PNchiral}
 \Pi_N =
 \left( \begin{array}{cc}
  C & C\mathcal{O}^t  \\
  \mathcal{O} C^t &  \mathcal{O} C^t \mathcal{O}^t
\end{array}\right) \ ,
\ee
where $\mathcal{O}$ is an $n \times n$ orthogonal matrix obeying
\be
 \mathcal{O} \mathcal{O}^t = \bid  \ ,
\ee
and $C$ is another $n \times n$ matrix that satisfies
\bea
 C_{n \times n} &=& \frac{1}{2} \left( \bid_{n \times n} + \tilde{C}_{n \times n} \right) \ , \label{C} \\
 \tilde{C} &=& - \tilde{C}^t \ ,  \qquad \tilde{C}^2 \,\, =\,\, -\tilde{C} \ , \label{Ct}
\eea
i.e., the off-diagonal part $\tilde{C}$ is antisymmetric. Once the solution \eqref{PNchiral} is fully determined, all possible Neumann projectors can be constructed from \eqref{PNchiral} by an $O(n,n)$ transformation.

Next, we show how \eqref{PNchiral} is derived.
Since the Neumann and Dirichlet projectors are related by $\bid = \Pi_N + \Pi^t_D$, we may write them as
\be \label{Pansatz}
 \Pi_N = \left( \begin{array}{cc}
  C & D  \\
  Q & R
\end{array}\right) \ , \qquad
\Pi^t_D = \left( \begin{array}{cc}
  \bid-C & -D  \\
  -Q &  \bid-R
\end{array}\right) \ ,
\ee
where $C,D,Q,R$ are all $n \times n$ matrices. With $\mathbb{H}$ and $\mathbb{L}$ given by \eqref{chiralLH}, the Neumann projector in the chiral frame is then obtained by solving, as a set, the projector condition \eqref{projc}, the null condition \eqref{nullc} and the orthogonality condition \eqref{orthocon}. Solving these conditions gives us the following constraints on the matrices,
\bea
C+C^t\!\!&\!\!=\!\!&\!\!\bid\ , \qquad C C^t=DD^t=\frac{C}{2} \ , \label{Peq1}\\
R+R^t\!\!&\!\!=\!\!&\!\!\bid \ , \qquad QQ^t=RR^t= \frac{R}{2} \ ,\label{Peq2}\\
D\!\!&\!\!=\!\!&\!\!Q^t \  \ . \label{Peq3}
\eea
It follows from \eqref{Peq1} that
\be \label{Peq4}
 C^2 = CC^t.
\ee
We then decompose $C$ into its symmetric and antisymmetric parts as in \eqref{C}. Since $C+C^t=\bid$, it is clear that the symmetric part is $\bid/2$ and that the antisymmetric part is the off-diagonal part. Applying \eqref{Peq4} to this decomposition we find $\tilde{C}^2 = -\tilde{C}$, as stated in \eqref{Ct}.

To solve for $D$, we start with the equation $CC^t = DD^t$ in \eqref{Peq1}.
The most general solution of $D$ is $D=C\mathcal{O}^t$, where $\mathcal{O}$ is an orthogonal matrix $\mathcal{O}\mathcal{O}^t =\bid$ so that $CC^t = DD^t = D\mathcal{O}\mathcal{O}^t D^t = D\mathcal{O} (D\mathcal{O})^t$. Similarly, $Q$ is obtained from \eqref{Peq3} as $Q=D^t=\mathcal{O}C^t$, while $R$ follows from \eqref{Peq2} resulting in $R=2D^t D = 2 \mathcal{O} C^t C \mathcal{O}^t = \mathcal{O} C^t \mathcal{O}^t$. Thus we have obtained the most general Neumann projector in the chiral frame, \eqref{PNchiral}.

We can transform the Neumann projector \eqref{PNchiral} from the chiral frame back to the original light-like coordinates using the vielbeins introduced in \eqref{vb}. The result is
\bea
\Pi_{N,I}{}^J &=& \mathcal{V}^t_I{}^{\underline{I}} \; \Pi_{N,\underline{I}}{}^{\underline{J}}\; \mathcal{V}^t_{\underline{J}}{}^J \nn\\
&=&\frac{1}{2}\left( \begin{array}{cc}
  \scriptstyle{ [(\bid+\mathbf{B})C-(\bid-\mathbf{B})\mathcal{O}C^t](\bid-\mathcal{O}^t)} & \scriptstyle{ [(\bid+\mathbf{B})C-(\bid-\mathbf{B})\mathcal{O}C^t][\bid-\mathbf{B}+\mathcal{O}^t (\bid+\mathbf{B})]} \\
  \scriptstyle{ (C+\mathcal{O} C^t) (\bid-\mathcal{O}^t)} & \scriptstyle{ (C+\mathcal{O} C^t)[\bid-\mathbf{B}+\mathcal{O}^t (\bid+\mathbf{B})]}
\end{array}\right) \ , \label{PNnull}
\eea
where $\mathbf{B}$ is the $n \times n$ $B$-field matrix, $\bid$ is the $n \times n$ unit matrix, $C$ is the upper-left $n \times n$ matrix  in \eqref{PNchiral} and $\mathcal{O}$ is the rotation matrix in the same expression.

Although the general form of $\Pi_N$ given above contains free parameters, other Neumann projectors can still be constructed from \eqref{PNchiral} and \eqref{PNnull} by $O(n,n)$ transformations. This is due to the fact that the three conditions --- the projector, null, and orthogonality conditions --- which are the basis for deriving the projectors, are $O(n,n)$ covariant. Thus other projectors $O(n,n)$-related to \eqref{PNchiral} (in the chiral frame) or to \eqref{PNnull} (in the light-like frame) can be obtained by solving the correspondingly $O(n,n)$-transformed triplet of conditions.

We have obtained the general form for the boundary projectors; in Appendix \ref{AppO22} we provide explicit examples for the $O(2,2)$ case.

\section{The $O(2,2)$ boundary projectors}
\label{AppO22}

In this section we explicitly derive the full set of  boundary Neumann and Dirichlet projectors in four-dimensional flat doubled space. We also demonstrate how to extract and interpret the physical meaning of each projector, labeling
each solution as Type I, Type II, etc.

Let the four-dimensional doubled space coordinates be denoted by
\be
\mathbb{X}=(X,Y,\tilde{X},\tilde{Y})^t \ .
\ee
We choose a flat target space metric $g_{ij}=\delta_{ij}$ and a constant background $B$-field $B_{ij}=B \epsilon_{ij}$ for $i,j=1,2$. Then the doubled metric $\mathbb{H}$ \eqref{Hmetric} and the $O(2,2)$ invariant metric $\mathbb{L}$ \eqref{L} become
\begin{equation} \label{HL22}
\mathbb{H}=\left( \begin{array}{cccc}
1+B^{2} & 0 & 0 & B  \\
0 & 1+B^2 & -B & 0\\
0 & -B & 1 & 0\\
B & 0 & 0 & 1
\end{array} \right) \ , \qquad
\mathbb{L}=\left( \begin{array}{cc}
0 & \bid_{2 \times 2}   \\
\bid_{2 \times 2} & 0
\end{array} \right) \ .
\end{equation}

Following the strategy described in Appendix \ref{AppOnn}, we find that the vielbeins necessary to transform between $\mathbb{R}^4$ and the chiral frame are
\be
\mathcal{V}^{\underline{I}}{}_I = \frac{1}{\sqrt{2}}
\left( \begin{array}{cccc}
  1 & -B & 1 & 0 \\
  B &  1 & 0 & 1 \\
 -1 & -B & 1 & 0 \\
  B & -1 & 0 & 1
\end{array}\right) \ , \qquad
\mathcal{V}^I{}_{\underline{I}} = \frac{1}{\sqrt{2}}
\left( \begin{array}{cccc}
  1 &  0 &-1 & 0 \\
  0 &  1 & 0 &-1 \\
  1 &  B & 1 &-B \\
 -B &  1 & B & 1
\end{array}\right) \ ,
\ee
which when applied to the metrics as in (\ref{vielbein}) yield the chiral frame form \eqref{chiralLH} of $\mathbb{H}$ and $\mathbb{L}$. The general Neumann and Dirichlet projectors in the chiral frame are given by \eqref{PNchiral}, but due to the property $\tilde{C}^2 = -\tilde{C}$, the antisymmetric off-diagonal part of $C$ vanishes, so that we are left with
\be \label{P22}
\Pi_N = \frac{1}{2}
\left( \begin{array}{cc}
  \bid_{2\times 2} &  \mathcal{O}^t_{2\times 2} \\
  \mathcal{O}_{2 \times 2} & \bid_{2\times 2}
\end{array}\right)\ , \qquad\quad
\Pi^t_D = \bid - \Pi_N \ .
\ee
All other chiral frame projectors are equivalent to \eqref{P22} up to some $O(n,n)$ transformations.

If we transform back to $\mathbb{R}^4$ using the vielbeins, the resulting Neumann projector reads
\be \label{PNO22}
\Pi_{N,I}{}^J = \mathcal{V}^t_I{}^{\underline{I}} \; \Pi_{N,\underline{I}}{}^{\underline{J}}\; \mathcal{V}^t_{\underline{J}}{}^J
=\frac{1}{2}\left[ \begin{array}{cc}
  (1-\cos\theta-B \sin\theta)\, \delta_{i}^{\ j} & [(B^2-1)\sin\theta + 2 B \cos\theta]\, \epsilon_{ij} \\
 (\sin\theta)\, \epsilon^{ij} & (1+\cos\theta+B \sin\theta)\, \delta^{i}_{\ j} \\
\end{array}\right]
\ ,
\ee
where $i,j=\{1,2\}$ and the $2 \times 2$ rotation matrix $\mathcal{O}$ is parameterized by a real number $\theta$,
\be
\mathcal{O} = \left( \begin{array}{cc}
  \cos \theta &  \sin \theta \\
  -\sin \theta & \cos \theta
\end{array}\right) \ .
\ee
Then one can construct other Neumann projectors in $\mathbb{R}^4$ from \eqref{PNO22} via $O(2,2)$ transformations.

On the other hand, the projectors can be obtained by directly solving for $4 \times 4$ real matrices in $\mathbb{R}^4$, satisfying the system of equations comprised by the projector condition \eqref{projc}, the null condition \eqref{nullc} and the orthogonality condition \eqref{orthocon}.
Regardless of whether we are working with sigma-type models \eqref{doubledaction} or  FJ-type models \eqref{FJ}, the boundary projectors must always satisfy these three conditions. Here we choose Hull's formalism \eqref{doubledaction}, which needs the self-duality constraint \eqref{selfdualcondition} to remove half of the doubled degrees of freedom. We will show how to obtain the boundary projectors by such a direct computation, but first let us write down the D-brane embedding implied by the Neumann projector  \eqref{PNO22}.

When expressed in component form, the self duality condition reads
\begin{equation}
\left( \begin{array}{c}
\partial_{0}X  \\
\partial_{0}Y \\
\partial_{0}\tilde{X}  \\
\partial_{0}\tilde{Y}
 \end{array}  \right)= \mathbb{L}\mathbb{H}\left( \begin{array}{c}
\partial_{1}X  \\
\partial_{1}Y \\
\partial_{1}\tilde{X}  \\
\partial_{1}\tilde{Y}
 \end{array}  \right)=
 \left( \begin{array}{c}
-B\partial_{1}Y+\partial_{1}\tilde{X}  \\
B\partial_{1}X+\partial_{1}\tilde{Y} \\
(1+B^{2})\partial_{1}X+B\partial_{1}\tilde{Y}  \\
(1+B^{2})\partial_{1}Y-B\partial_{1}\tilde{X}
 \end{array}  \right),
\end{equation}
where $\partial_{0}\equiv\partial_{\sigma^0}$ and $\partial_{1}\equiv\partial_{\sigma^1}$. From this set of conditions immediately follows the relations
\begin{equation}
\label{selfdualexamp}
\begin{array}{rcl}
\partial_{0}\tilde{X} &=& \partial_{1}X+B\partial_{0}Y\,,\\
\partial_{0}\tilde{Y} &=& \partial_{1}Y-B\partial_{0}X \,,\\
\partial_{1}\tilde{X} &=& \partial_{0}X+B\partial_{1}Y \,,\\
\partial_{1}\tilde{Y} &=& \partial_{0}Y-B\partial_{1}X \,.
\end{array}
\end{equation}
Using these relations to eliminate the dual coordinates $\{\tilde{X},\tilde{Y}\}$ in the Neumann boundary condition (\ref{Ncond2}),
$\Pi_N \, \mathbb{H}\, \partial_1 \mathbb{X} = 0$, with $\Pi_N$ given by \eqref{PNO22}, we find
\begin{equation}
\label{thetaparambc}
\begin{array}{rcl}
 (1+\cos \theta) \partial_0 Y - \sin \theta\,\partial_1 X &=& 0 \,, \\
 (1+\cos \theta) \partial_0 X + \sin \theta\,\partial_1 Y &=& 0 \,,
\end{array}
\end{equation}
i.e., a pair of boundary conditions for the two physical coordinates $\{X,Y\}$, which dictate how a D-brane may be embedded in the physical subspace. This pair of boundary conditions corresponds to  D0- or D2-branes,
depending on the choice of the parameter $\theta$.

Below we derive the $O(2,2)$ boundary projectors by direct computation in $\mathbb{R}^4$, and show that we again obtain the
boundary conditions \eqref{thetaparambc}, plus a few D1-brane conditions. We
provide a physical interpretation for each of the D-brane solutions, which we label Type I-VIII.

From the projector condition \eqref{projc} and the null condition \eqref{nullc} we find that the Dirichlet projector must take the form
\begin{equation}
\Pi_{D}=\left[ \begin{array}{cc}
 \tilde{a} & \tilde{b}  \\
 \tilde{c} & \bid_{2\times 2}-\tilde{a}^{t}  \\
 \end{array}  \right] \,,
\end{equation}
where $\tilde{a}$, $\tilde{b}$ and $\tilde{c}$ are $2\times 2$ matrices satisfying
\begin{eqnarray}
&& \tilde{b}^{t}=-\tilde{b}, \quad \tilde{c}^{t}=-\tilde{c},\quad \tilde{b}\tilde{c}=\tilde{a}(\bid_{2\times 2}-\tilde{a}),\nonumber\\
&& \tilde{a}\tilde{b}=-(\tilde{a}\tilde{b})^{t},\quad \tilde{c}\tilde{a}=-(\tilde{c}\tilde{a})^{t},
\end{eqnarray}
and the superscript $t$ stands for transpose.
After imposing also the orthogonality condition \eqref{orthocon}, we find the following types of solutions.

\textbf{Type I:}
\begin{equation}
\Pi_{D}=\left(\begin{array}{cccc}
a_{1} & 0 & 0 & c_{1} \\
0 & a_{1} & -c_{1} & 0  \\
0 & b_{1} & 1-a_{1} & 0\\
-b_{1} & 0 & 0 & 1-a_{1}
 \end{array}\right),  \qquad
\Pi_{N}=\left(\begin{array}{cccc}
1-a_{1} & 0 & 0 & b_{1} \\
0 & 1-a_{1} & -b_{1} & 0  \\
0 & c_{1} & a_{1} & 0\\
-c_{1} & 0 & 0 & a_{1}
 \end{array}\right),
\end{equation}
where $a_{1}$, $b_{1}$ and $c_{1}$ are real constants satisfying
\be
\label{abcconds1}
\begin{cases}
& a_{1}^{2}-b_{1}c_{1}-a_{1}=0 \ , \\
& c_{1}^{2}+c_{1}^{2} B^{2}-2 \ c_{1}a_{1} \, B + c_1\,B +a_{1}^{2} - a_{1}=0 \ , \\
& a_1 \neq \{0,1\}, \quad b_1 \neq 0, \quad c_1 \neq 0 \ ,
\end{cases}
\ee
for $B \neq 0$.
Inserting the Type I Neumann projector together with the self-duality relations \eqref{selfdualexamp} into the Neumann boundary condition $\Pi_N \, \mathbb{H} \,\partial_1 \mathbb{X}=0$ yields the boundary conditions expressed in terms of the physical coordinates,
\begin{equation}
\label{TypeIbc}
\begin{cases}
& \partial_1 X + \left( B + \frac{b_1}{1-a_1}\right) \partial_0 Y = 0  \ , \\
& \partial_1 Y - \left( B + \frac{b_1}{1-a_1}\right) \partial_0 X = 0  \ ,
\end{cases}
\end{equation}
where $B+b_1/(1-a_1)\neq 0$ due to \eqref{abcconds1}.
Note that the boundary conditions \eqref{TypeIbc} are expressed entirely in the context of the two-dimensional physical space --- we have used the self-duality constraint to leave the doubled domain by eliminating the dual coordinates, so that \eqref{TypeIbc} corresponds to the actual physical D-branes. So what kind of D-branes are they?

For intermediate values of $B$ and the parameters $a_1, b_1, c_1$, the boundary conditions resemble the ordinary Neumann conditions for a D2-brane that couples to a background $B$-field. However, there is an extra contribution $b_1/(1-a_1)$ which appears to augment the coupling to $B$, and this is in effect what happens. Let us analyze the configuration in the same manner as in \cite{Albertsson:2008gq}, section 4.1.3.
First consider the doubled space Dirichlet conditions for the case at hand,
\begin{equation}
\label{TypeIDc}
\begin{cases}
& \partial_0 \tilde{X} + \frac{b_1}{1-a_1} \partial_0 Y = 0  \ , \\
& \partial_0 \tilde{Y} - \frac{b_1}{1-a_1} \partial_0 X = 0  \ .
\end{cases}
\end{equation}
This brane (which is always two-dimensional from the doubled perspective)
is a straight line in the $\{X,\tilde{Y}\}$ plane with inclination $b_1/(1-a_1)$ and
a straight line in the $\{Y,\tilde{X}\}$ plane with inclination $-b_1/(1-a_1)$. For small values of
the inclination, the brane all but coincides with the physical space $\{X,Y\}$, so that \eqref{TypeIDc}
approaches a pair of Dirichlet conditions for $\tilde{X}$ and $\tilde{Y}$, and the physical
conditions \eqref{TypeIbc} reduce to normal Neumann conditions with coupling to $B$. For large values of
the inclination on the other hand, the brane lies almost entirely in the $\{\tilde{X},\tilde{Y}\}$ plane, which
from the point of view of the $\{X,Y\}$ plane means that the motion of a string on the brane appears
 restricted,
a situation that suggests the presence of a strong field.
In \eqref{TypeIDc} this is represented by the $\partial_0 X$ and
$\partial_0 Y$ parts becoming dominant in such a way that the two equations approach
Dirichlet conditions for $X$ and $Y$ at the same pace. The same asymptote is manifest
in the physical space conditions \eqref{TypeIbc}, except here the $B$-field from the doubled
metric enters too.

It may appear strange to be left with an artifact from the doubled formalism, in the form of
the ``extra $B$-field'' $b_1/(1-a_1)$, but recall that the field $B$ that appears in the doubled metric
does not necessarily coincide with the $B$-field in physical space after we have projected
our theory down from the doubled space \cite{Hull:2009sg,ReidEdwards:2009nu,Dall'Agata:2007sr}.
The final physical $B$-field is a combination of the
doubled geometry properties (i.e., the component $B$ of the doubled metric) and the orientation
of double D-branes (here parameterized by the inclination $b_1/(1-a_1)$. 

Finally, we note that this solution is identical to the one in \eqref{PNO22}, if we define
\be
 a_1 = \frac{1}{2}\,(1 + \cos \theta + B\, \sin \theta)\ , \qquad
 b_1 = B\, \cos \theta + \frac{B^2 - 1}{2}\, \sin \theta \ , \qquad
 c_1 = \frac{1}{2}\,\sin \theta\ .
\ee

\textbf{Type II:}
\begin{equation}
\Pi_{D}=\left(\begin{array}{cccc}
0 & 0 & 0 & \frac{-B}{1+B^2} \\
0 & 0 & \frac{B}{1+B^2} & 0  \\
0 & 0 & 1 & 0   \\
0 & 0 & 0 & 1
 \end{array}\right), \qquad
\Pi_{N}=\left(\begin{array}{cccc}
1 & 0 & 0 & 0 \\
0 & 1 & 0 & 0  \\
0 & \frac{-B}{1+B^2} & 0 & 0   \\
\frac{B}{1+B^2} & 0 & 0 & 0
 \end{array}\right) \,.
\end{equation}
for $B\neq 0$.
The Neumann boundary condition reduced by the self-duality constraint amounts to
\be
\begin{cases}
 \partial_1 X + B \partial_0 Y = 0 \,, \\
 \partial_1 Y -B \partial_0 X = 0 .
\end{cases}
\ee
This is a D2-brane spanning the $\{X,Y\}$ directions and coupled to a constant $B$-field in the physical space.
This type of projector is identical to \eqref{PNO22} under the mapping
\be
\sin\theta = \frac{-2B}{B^2+1}\ , \qquad \cos \theta = \frac{B^2 -1}{B^2+1} \ .
\ee

\textbf{Type III:}
\begin{equation}
\Pi_{D}=\left(\begin{array}{cccc}
1 & 0 & 0 & \frac{B}{1+B^2}  \\
0 & 1 & \frac{-B}{1+B^2} & 0  \\
0 & 0 & 0 & 0\\
0 & 0 & 0 & 0
 \end{array}\right), \qquad
\Pi_{N}=\left(\begin{array}{cccc}
0 & 0 & 0 & 0  \\
0 & 0 & 0 & 0  \\
0 & \frac{B}{1+B^2} & 1 & 0\\
\frac{-B}{1+B^2} & 0 & 0 & 1
 \end{array}\right),
\end{equation}
for all $B$. The corresponding boundary conditions are
\be
\label{TypeIIIbc}
\begin{cases}
& B \partial_1 Y + \partial_0 X = 0 \ , \\
& B \partial_1 X -  \partial_0 Y = 0 \ .
\end{cases}
\ee
For $B \neq 0$ this is a D2-brane coupled to a $B$-field $1/B$ in the physical space, and we again see how the projection from doubled space can produce a physical $B$-field different from the one in the doubled metric $\mathbb{H}$. For $B=0$ \eqref{TypeIIIbc} describes a $D0$-brane. 
The Type III projector is identical to \eqref{PNO22} if we set
\be
\sin\theta = \frac{2B}{1+B^2}\ , \qquad \cos \theta = \frac{1-B^2}{1+B^2} \ .
\ee

\textbf{Type IV:} 
\begin{equation}
\Pi_{D}=\left(\begin{array}{cccc}
0 & 0 & 0 & 0 \\
0 & 0 & 0 & 0  \\
0 & -B & 1 & 0   \\
B & 0 & 0 & 1
 \end{array}\right), \qquad
\Pi_{N}=\left(\begin{array}{cccc}
1 & 0 & 0 & -B \\
0 & 1 & B & 0  \\
0 & 0 & 0 & 0   \\
0 & 0 & 0 & 0
 \end{array}\right) \,,
\end{equation}
for all $B$.
The Neumann boundary condition together with the self-duality constraint yield
\be
 \partial_1 X  =  \partial_1 Y  = 0 \ .
\ee
These boundary conditions describe a D2-brane without coupling to a background $B$-field even when $B \neq 0$. The doubled space Dirichlet conditions read
\begin{equation}
\label{TypeIVDc}
\begin{cases}
& \partial_0 \tilde{X} -B \partial_0 Y = 0  \ , \\
& \partial_0 \tilde{Y} +B \partial_0 X = 0  \ ,
\end{cases}
\end{equation}
whence we see that $B$ determines the orientation of the D-brane in four-dimensional
space. However, the same $B$-field appears in the self-duality constraint \eqref{selfdualexamp} in such a
 way as to completely cancel its effect on the two-dimensional theory after projecting
 down to physical space, and we end up without a physical $B$-field coupling.
This type of projector is seen to coincide with \eqref{PNO22} by setting
\be
\sin\theta = 0\ , \qquad \cos \theta = -1 \ .
\ee

\textbf{Type V:}
\begin{equation}
\Pi_{D}=\left(\begin{array}{cccc}
1 & 0 & 0 & 0 \\
0 & 1 & 0 & 0  \\
0 & B & 0 & 0   \\
-B & 0 & 0 & 0
 \end{array}\right), \qquad
\Pi_{N}=\left(\begin{array}{cccc}
0 & 0 & 0 & B \\
0 & 0 & -B & 0  \\
0 & 0 & 1 & 0   \\
0 & 0 & 0 & 1
\end{array}\right)\,,
\end{equation}
for all $B$.
The reduced boundary condition reads
\be
 \partial_0 X  = \partial_0 Y = 0 \ .
\ee
These boundary conditions describe a D0-brane in the physical space.
This type of projector is identical to \eqref{PNO22} if we set
\be
\sin\theta = 0\ , \qquad \cos \theta = 1 \ .
\ee

\textbf{Type VI:}
\begin{equation}
\label{TypeVIproj}
\Pi_{D}=\left(\begin{array}{cccc}
0 & 0 & 0 & 0  \\
0 & 1 & 0 & 0  \\
0 & 0 & 1 & 0\\
0 & 0 & 0 & 0
 \end{array}\right),\qquad
\Pi_{N}=\left(\begin{array}{cccc}
1 & 0 & 0 & 0  \\
0 & 0 & 0 & 0  \\
0 & 0 & 0 & 0\\
0 & 0 & 0 & 1
 \end{array}\right).
\end{equation}
These projectors correspond to the boundary conditions
\begin{equation}
 \partial_{0} Y = \partial_{1} X = 0 \ .
\end{equation}
Using the self-duality constraint, one finds that the double D-brane intersects the $O(2,2)$ double space in the  $\{X,\tilde{Y}\}$-coordinates, and thus it is a D1-brane spanning the physical dimension $\{X\}$.

Here we cannot reproduce the projector \eqref{PNO22} simply by choosing values for $\theta$;
we need to first apply an $O(2,2)$ transformation either to \eqref{PNO22} or to \eqref{TypeVIproj},
so as to turn the two solutions into branes of the same dimension.

\textbf{Type VII:}
\begin{equation}
\Pi_{D}=\left(\begin{array}{cccc}
1 & 0 & 0 & 0  \\
0 & 0 & 0 & 0  \\
0 & 0 & 0 & 0 \\
0 & 0 & 0 & 1
\end{array}\right), \qquad
\Pi_{N}=\left(\begin{array}{cccc}
0 & 0 & 0 & 0  \\
0 & 1 & 0 & 0  \\
0 & 0 & 1 & 0 \\
0 & 0 & 0 & 0
\end{array}\right),
\end{equation}
with corresponding boundary conditions
\begin{equation}
 \partial_{0} X = \partial_1 Y= 0  \ .
\end{equation}
This case is analogous to the Type VI solution. The worldvolume of this brane coincides with the $\{Y,\tilde{X}\}$-plane, so we have a D1-brane along the physical direction \{Y\}. Equivalence to \eqref{PNO22} can again be shown by first performing an $O(2,2)$ transformation.

\textbf{Type VIII:}
\begin{equation}
\Pi_{D}=\left(\begin{array}{cccc}
1-a_{4} & a_{2} & 0 & 0  \\
\frac{a_{4}(1-a_{4})}{a_{2}} & a_{4} & 0 & 0  \\
0 & 0 & a_{4} & -\frac{a_{4}(1-a_{4})}{a_{2}}\\
0 & 0 & -a_{2} & 1-a_{4}
 \end{array}\right), \qquad
\Pi_{N}=\left(\begin{array}{cccc}
a_{4} & -\frac{a_{4}(1-a_{4})}{a_{2}} & 0 & 0  \\
-a_2 & 1- a_{4} & 0 & 0  \\
0 & 0 & 1-a_{4} & a_2\\
0 & 0 & \frac{a_{4}(1-a_{4})}{a_{2}} & a_{4}
 \end{array}\right),
\end{equation}
where $a_{2}$ and $a_{4}$ are constants constrained by
\begin{equation}
a_{2}^{2}=a_{4}(1-a_{4}) \ , \qquad a_4 \neq \{ 0,1 \}\ ,\  a_2 \neq 0 \ ,
\end{equation}
and the corresponding boundary condition in terms of the physical coordinates is
\begin{equation}
\begin{cases}
& \partial_{0} X + \frac{a_{2}}{1-a_{4}}\, \partial_{0} Y=0, \\
& \partial_{1} Y - \frac{a_{4}}{a_{2}}\, \partial_{1} X=0.
\end{cases} \ ,
\end{equation}
This configuration is a $D1$-brane lying in the $\{X,Y\}$ plane, and it is independent of $B$.
A suitable combination of a $O(2,2)$ transformation and a $\theta$-choice again reproduces
the projector \eqref{PNO22}.

\section{Calculation details of the Neumann Green's functions}
\label{AppGreen}

In this section we derive the Green's function in the $O(n,n)$ chiral frame.
The Neumann Green's functions $G^{\underline{I}\underline{J}}$ and $G^{TT}$ are governed by the following equations of motion,
\bea
 \mathbb{H}_{\underline{K}\underline{I}}\, \partial_1^2 G^{\underline{I}\underline{J}} -\mathrm{i}\, \mathbb{L}_{\underline{K}\underline{I}} \partial_0 \partial_1 G^{\underline{I}\underline{J}} &=& - \delta_{\underline{K}}{}^{\underline{J}} \, 2 \pi\, \delta(\vec{\sigma}-\vec{\sigma}') \ , \label{eomG} \\
\delta^{\alpha\beta} \partial_{\alpha}\partial_{\beta} G^{TT} &=&  - 2\pi \delta(\vec{\sigma}-\vec{\sigma}') \ . \label{eomTG}
\eea
These two equations can be combined to give (\ref{Oin1eom}) by introducing $\hat{\mathbb{H}}$ and $\hat{\mathbb{L}}$ as in (\ref{Oin1G}). (\ref{eomG}) can be expressed explicitly in the chiral frame in terms of the complexified worldsheet coordinates $z = \sigma + \mathrm{i}\tau$, $\bar{z}=\sigma - \mathrm{i}\tau$, as follows \cite{Berman:2007xn},
\be
 \left(\begin{array}{cc}
  \delta_{\underline{i}\underline{j}} (\partial_{z}+\partial_{\bar{z}})\partial_z G^{\underline{j}\underline{k}} & 0 \\
0 &  \delta_{\underline{i}'\underline{j}'} (\partial_{z}+\partial_{\bar{z}})\partial_{\bar{z}} G^{\underline{j}'\underline{k}'}
 \end{array}\right)
 = \left(\begin{array}{cc}
 \delta_{\underline{i}}{}^{\underline{k}} \pi \delta(z-z') & 0 \\
 0 & \delta_{\underline{i}'}{}^{\underline{k}'} \pi \delta(z-z')
 \end{array}\right) \ .
\ee
This yields the two equations
\bea
 \delta_{\underline{i}\underline{j}} (\partial_{z}+\partial_{\bar{z}})\partial_z G^{\underline{j}\underline{k}} &=&
 \frac{1}{2}\delta_{\underline{i}\underline{j}} \partial_{1}\partial_- G_+^{\underline{j}\underline{k}} \ = \
  \delta_{\underline{i}}{}^{\underline{k}} \pi \delta(z-z') \ , \nn \\
\delta_{\underline{i}'\underline{j}'} (\partial_{z}+\partial_{\bar{z}})\partial_{\bar{z}} G^{\underline{j}'\underline{k}'} &=&
 \frac{1}{2}\delta_{\underline{i}'\underline{j}'} \partial_{1}\partial_+ G_-^{\underline{j}'\underline{k}'} \ = \
 \delta_{\underline{i}'}{}^{\underline{k}'} \pi \delta(z-z') \ , \nn
\eea
which determine the Green's functions for the left and right chiral scalars associated with the FJ action.
Then $G_+^{\underline{j}\underline{k}}$ and $G_-^{\underline{j}'\underline{k}'}$ are obtained as
\bea
 G_+^{\underline{j}\underline{k}} &=&
 -\frac{\delta^{\underline{j}\underline{k}}}{2\pi} \ln (\bar{z}-\bar{z}') = \delta^{\underline{j}\underline{k}} \Delta_+\ , \qquad \Delta_+ = -\frac{1}{2\pi} \ln (\bar{z}-\bar{z}') \ ,\nn \\
 G_-^{\underline{j}'\underline{k}'} &=&
 -\frac{\delta^{\underline{j}'\underline{k}'}}{2\pi} \ln(z-z')
 = \delta^{\underline{j}'\underline{k}'} \Delta_- \ , \qquad
 \Delta_- = -\frac{1}{2\pi} \ln(z-z')\ , \nn
\eea
so that the full Green's function $G^{\underline{J}\underline{K}}$ reads
\be\label{finalG}
G^{\underline{J}\underline{K}} = G_+^{\underline{j}\underline{k}} + G_-^{\underline{j}'\underline{k}'} = \frac{1}{2}(\mathbb{H}+\mathbb{L})^{\underline{J}\underline{K}}\Delta_+ +\frac{1}{2} (\mathbb{H}-\mathbb{L})^{\underline{J}\underline{K}}\Delta_-
= \mathbb{H}^{\underline{J}\underline{K}}\; \Delta_0 + \mathbb{L}^{\underline{J}\underline{K}} \; \Theta \ ,
\ee
where
\be\label{bulkGr}
 \Delta_0 = \frac{1}{2} (\Delta_+ + \Delta_-)\ , \qquad
 \Theta = \frac{1}{2} (\Delta_+ - \Delta_-)\;.
\ee
The propagators $\Delta_{\pm}$ are two-dimensional propagators for the left and right chiral scalars, respectively \cite{Tseytlin:1990va,Berman:2007xn}.

The Green's function $G^{\underline{J}\underline{K}}$ was obtained without imposing any boundary conditions. Note that in (\ref{finalG}) the Green's function is $O(n,n)$ covariant, hence the solution holds in any $O(n,n)$ frame. Moreover,  projecting to the Neumann subspace, we obtain the Neumann part of the Green's function,
\be\label{bndyGr}
G^{\underline{p}\underline{q}}  = \mathcal{H}^{\underline{p}\underline{q}}\; \Delta_0 \,.
\ee
Note that the $\Theta$ term in (\ref{bulkGr}) is absent in (\ref{bndyGr}) due to the null condition (\ref{nullc}), i.e., $\Pi_N^t \mathbb{L}^{-1} \Pi_N=0$.

Combining the Neumann part of the Green's function in doubled space with the trivial solution for $G^{TT}$, i.e., $G^{TT}=- \frac{1}{4\pi} (\ln (z-z') + \ln (\bar{z}-\bar{z}'))$, one obtains the ansatz (\ref{Gansatz}). Substituting (\ref{Gansatz}) into the boundary condition (\ref{Oin1bc}), it is then straightforward to find the mirror image part $\mathbf{u}$ on the boundary.

Note that in \cite{Berman:2007xn}, the Wick rotation is chosen to be $\mathrm{i} \tau_E = t_M$ while we adopt the convention with an opposite sign, thus their $\partial_{t_E}$ expressed in complex coordinates has an opposite sign compared to ours, and it follows that $\Delta_+$ and $\Delta_-$ in this paper have a different sign convention compared to that of \cite{Berman:2007xn}.

\end{document}